%% file: emratio.tex
\documentclass[aps,prc,nofootinbib,groupedaddress,twocolumn,10pt,floatfix]{revtex4-2}

\usepackage{emratio}

\ifproofpre{}{\input{vc/vc.tex}}

\begin{document}


\title{Quadrupole moments and proton-neutron structure in $p$-shell mirror nuclei}

\author{Mark A.~Caprio}
\affiliation{Department of Physics, University of Notre Dame, Notre Dame, Indiana 46556-5670, USA}

\author{Patrick J.~Fasano}
\affiliation{Department of Physics, University of Notre Dame, Notre Dame, Indiana 46556-5670, USA}

\author{Pieter Maris}
\affiliation{Department of Physics and Astronomy, Iowa State University, Ames, Iowa 50011-3160, USA}

\author{Anna E. McCoy}
\affiliation{Institute for Nuclear Theory, University of Washington, Seattle, Washington 98195-1550, USA}

\date{\today}

\begin{abstract}
  \input{emratio-abstract}
\end{abstract}

\preprint{Git hash: \VCRevisionMod}

\maketitle


\input{emratio-intro}
\raggedbottom  
\input{emratio-nuclei}
\input{emratio-ratio}
\flushbottom
\input{emratio-symm}
\input{emratio-discussion}

\input{emratio-concl}




\begin{acknowledgments}
We thank Saori Pastore and James P.~Vary for valuable
discussions and Jingru Z.~Benner, Colin V.~Coane, Jakub Herko, and Zhou Zhou for
comments on the paper.  This material is based upon work supported by the
U.S.~Department of Energy, Office of Science, under Award Numbers
DE-FG02-95ER40934, DE-FG02-00ER41132,  and DESC00018223 (SciDAC4/NUCLEI).  An award of computer time
was provided by the Innovative and Novel Computational Impact on Theory and
Experiment (INCITE) program. This research used computational resources of the
National Energy Research Scientific Computing Center (NERSC) and the Argonne
Leadership Computing Facility (ALCF), which are U.S.~Department of Energy,
Office of Science, user facilities, supported under Contract Numbers
DE-AC02-05CH11231 and DE-AC02-06CH11357.
\end{acknowledgments}


\nocite{control:title-on}
\input{emratio.bbl}



\end{document}

%% file: emratio-abstract.tex
%
%
%
%
Electric quadrupole ($E2$) matrix elements provide a measure of nuclear
deformation and related collective structure.  Ground-state quadrupole moments
in particular are known to high precision in many $p$-shell nuclei.
%
%
%
%
While the experimental electric quadrupole moment only measures the proton
distribution, both proton and neutron quadrupole moments are needed to probe
proton-neutron asymmetry in the nuclear deformation.
%
%
%
%
We seek insight into the relation between these moments through the \textit{ab
initio} no-core configuration interaction (NCCI), or no-core shell model (NCSM),
approach.  Converged \textit{ab initio} calculations for quadrupole moments are
particularly challenging, due to sensitivity to long-range behavior of the wave
functions.  We therefore study more robustly-converged \textit{ratios} of
quadrupole moments: across mirror nuclides, or of proton and neutron quadrupole
moments within the same nuclide.
%
%
%
%
In calculations for mirror pairs in the $p$ shell, we explore how well the
predictions for mirror quadrupole moments agree with experiment and how well
isospin (mirror) symmetry holds for quadrupole moments across a mirror pair.
%
%
%
%
%
The comparison with experiment confirms the predictive power of the \textit{ab
initio} description, indicating that the predicted ratios are physically
relevant for understanding proton-neutron structure as well.
%
%
%
%
%
%
%
%
\relax

%% file: emratio-intro.tex
\section{Introduction}
\label{sec:intro}

Electric quadrupole ($E2$) matrix elements, including quadrupole moments,
provide a principal measure of nuclear deformation, rotation, and related
collective
structure~\cite{eisenberg1987:v1,bohr1998:v2,casten2000:ns,rowe2010:collective-motion,rowe2010:rowanwood}.
To probe the proton-neutron asymmetric aspects of the nuclear deformation, both
proton and neutron quadrupole observables are needed.  For instance, in an
axially symmetric rotational nucleus~\cite{rowe2010:collective-motion}, the
ratio of proton and neutron quadrupole moments indicates the relative
contributions of protons and neutrons to the overall deformation, that is, the
ratio of quadrupole moments in the rotational intrinsic frame.  Alternatively,
in a spherical shell-model description, an anomalously large quadrupole moment for one of the
nucleonic species can be interpreted as an indicator of halo
structure~\cite{minamisono1992:8b-quadrupole-beta-nmr,kitagawa1993:shell-8b-17f-quadrupole-moment-halo}.

While the ground state electric quadrupole moment is readily accessible via
electromagnetic measurements~\cite{stone2016:e2-moments}, this observable
provides access only to the quadrupole moment of the proton density distribution
within the nucleus.  Neutron quadrupole observables are at best indirectly
measurable, \textit{e.g.}, through nuclear inelastic hadron or $\alpha$
scattering~\cite{bernstein1981:pn-me-hadron-scatt,*bernstein1983:pn-me-hadron-scatt}.
For the neutron quadrupole moment, in particular, if we consider two nuclei
forming a mirror pair, the approximate isospin symmetry of the nuclear
system~\cite{henley1969:isospin-nuclear-forces} implies that the behavior of the
protons in one member of the pair provides a proxy for the behavior of the
neutrons in the other, and \textit{vice versa}.  Thus, by mirror symmetry, we
can deduce the neutron quadrupole moment of one nucleus from the measured proton
quadrupole moment of the other.  However, this approach is limited to cases in
which the quadrupole moments are experimentally accessible for both members of
the mirror pair.  Moreover, it relies on the assumption, possibly imperfect, of
mirror symmetry.

We therefore seek further insight from \textit{ab initio} theory into the proton
and neutron quadrupole moments, and their relation, in light ($p$-shell) nuclei.
\textit{Ab initio} approaches to the nuclear many-body problem do not impose any
specific \textit{a priori} model assumptions, yet they are found to reproduce
signatures of phenomena involving quadrupole deformation,
clustering~\cite{pieper2004:gfmc-a6-8,neff2004:cluster-fmd,maris2012:mfdn-ccp11,yoshida2013:ncmcsm-8be-10be-6be-cluster,romeroredondo2016:6he-correlations,navratil2016:ncsmc},
and
rotation~\cite{caprio2013:berotor,maris2015:berotor2,*maris2019:berotor2-ERRATUM,caprio2015:berotor-ijmpe,stroberg2016:ab-initio-sd-multireference,jansen2016:sd-shell-ab-initio,caprio2020:bebands}.

Before extracting physical information from \textit{ab initio} calculations, we
must first identify the extent to which convergence is obtained for the relevant
observables.  In practice, the many-body Hilbert space used in an \textit{ab
  initio} calculation must be truncated to finite size.  Calculations in
progressively larger spaces, involving progressively less severe truncations,
provide results which converge towards those which would be obtained by solving
the full, untruncated many-body problem.  Nonetheless, the accuracy of
calculated observables is limited by computational constraints on the spaces
which can be accommodated.

The degree of convergence varies tremendously depending upon the observable
under consideration.  It is particularly challenging to obtain meaningful
\textit{ab initio} calculations for $E2$ matrix elements, including quadrupole
moments, due to their sensitivity to the large-distance behavior (or tails) of the
wave functions.  While the no-core configuration interaction
(NCCI)~\cite{barrett2013:ncsm}, or no-core shell model (NCSM), approach has
considerable success in providing calculations for, \textit{e.g.}, energies and
magnetic dipole observables, throughout the $p$
shell~\cite{maris2013:ncsm-pshell}, even order-of-magnitude calculations for
$E2$ matrix elements are elusive.

To circumvent this limitation, we recognize that \textit{ratios} of $E2$ matrix
elements can be more robustly calculated than individual matrix elements.  If
the matrix elements entering into the ratio involve structurally similar states,
so that convergence properties are similar, errors from truncation of the
many-body space can cancel in the ratio.  Such behavior has already been noted
for $E2$ matrix elements among states within the same rotational
band~\cite{maris2015:berotor2,caprio2020:bebands}, or between rotational bands
with related structures~\cite{caprio2019:bebands-sdanca19}, and for the $E2$
strengths of mirror transitions~\cite{henderson2019:7be-coulex}.  Prior
calculations of the ratio of proton and neutron quadrupole
moments~\cite{maris2015:berotor2} have suggested that this ratio may be
similarly robust, at least in a limited sampling of rotational states (see
Fig.~18 of Ref.~\cite{maris2015:berotor2}).

In the present work, to investigate proton-neutron asymmetry in quadrupole
structure, we consider NCCI calculations for two ratios: (1)~that of the proton
(or electric) quadrupole moments for the ground states of both members of a
mirror pair, as accessible in experiment, and (2)~that of the proton and neutron
quadrupole moments within the ground state of a single nuclide, which provides
the more direct structural measure.  To the extent that convergence is obtained
for these ratios, and to the extent that the results are found to be robust with
respect to the choice of internucleon interaction, we wish to understand the
following:

(1) How well do the predictions agree with experiment, in mirror pairs for which
the quadrupole moments of both members are experimentally
known~\cite{stone2016:e2-moments}?

(2) To what extent does mirror symmetry actually hold for the proton and neutron
quadrupole moments across a mirror pair, so that the mirror ratio may indeed be
taken as equivalent to the ratio of proton and neutron quadrupole moments within
a single member of the pair?

We carry out NCCI calculations of ground-state quadrupole moments for a
comprehensive set of pairs of mirror nuclei in the $p$ shell.  Namely, we
consider all mirror pairs in which both members are particle-bound (as well as
one involving an extremely narrow ground-state resonance) and in which the angular momentum selection rule permits a
nonzero quadrupole moment.  We compare results based on the
Daejeon16 interaction~\cite{shirokov2016:nn-daejeon16},
JISP16 interaction~\cite{shirokov2007:nn-jisp16}, and LENPIC interaction taken to \ntwolo{} in chiral effective
field theory (EFT)~\cite{epelbaum2015:lenpic-n4lo-scs,epelbaum2015:lenpic-n3lo-scs}.

In the following, after a brief overview of the mirror pairs being considered
and the available experimental data (Sec.~\ref{sec:nuclei}), we examine the
convergence properties of the ground-state quadrupole moments, and their ratios,
obtained in the NCCI calculations for these nuclei (Sec.~\ref{sec:ratio}).  We
then examine the accuracy with which mirror symmetry holds for the calculated
ratios, given the isospin symmetry breaking provided by the interactions
considered here (Sec.~\ref{sec:symm}).  Approaches to interpreting the
calculated ratio of neutron and quadrupole moments as an indicator of
proton-neutron nuclear structure, in terms of
cluster~\cite{freer2007:cluster-structures} or
$\grpsu{3}$~\cite{elliott1958:su3-part1,*elliott1958:su3-part2,*elliott1963:su3-part3,*elliott1968:su3-part4,harvey1968:su3-shell}
descriptions, are suggested (Sec.~\ref{sec:discussion}).

%% file: emratio-nuclei.tex
\section{\boldmath Mirror quadrupole moments in the $p$ shell}
\label{sec:nuclei}
\begin{figure}
\begin{center}
\includegraphics[width=\ifproofpre{1}{1}\hsize]{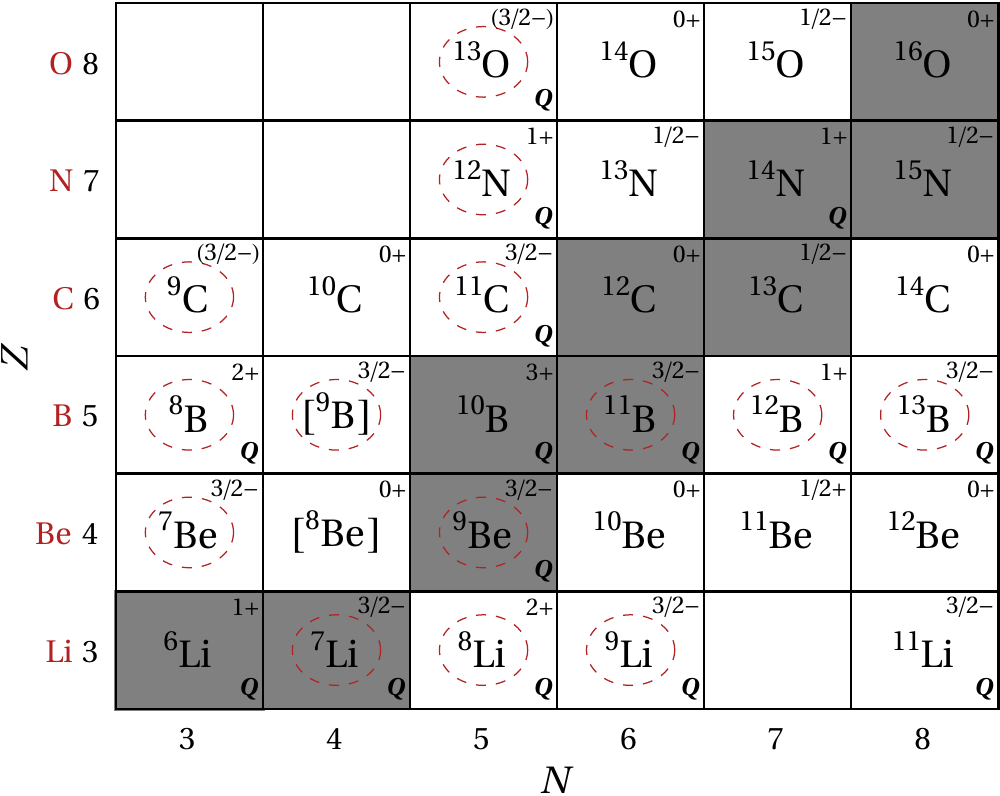}
\end{center}
\caption{Overview of particle-bound nuclides in the $p$ shell, indicating ground-state
  angular momentum and parity~\cite{npa2002:005-007,npa2004:008-010,npa1990:011-012,npa1991:013-015,npa2012:011}.
  Brackets indicate a particle-unbound but narrow ($\lesssim 1\,\keV$)
  ground-state resonance.  Nuclides
  with measured ground-state quadrupole moments~\cite{stone2016:e2-moments} are indicated with the letter
  ``$Q$''.  Those cases in which both members of a
  mirror pair are particle-bound and have ground-state angular momenta
  supporting a quadrupole moment, the criterion for inclusion in the present
  calculations, are highlighted (dashed circles). Shading indicates stable isotopes.
\label{fig:nuclear-chart}
}
\end{figure}
\input{emratio_tab02}

We focus on mirror pairs in the $p$ shell for which both members are
particle-bound.  The full set of bound $p$-shell
nuclei~\cite{npa2002:005-007,npa2004:008-010,npa1990:011-012,npa1991:013-015,npa2012:011}
is summarized in Fig.~\ref{fig:nuclear-chart}, along with ground-state angular
momentum and parity assignments.  For present purposes, we also include
exceptionally narrow ground-states resonances, indicated by brackets in
Fig.~\ref{fig:nuclear-chart}. (In particular, although $\isotope[9]{B}$ is
unstable to proton emission, its ground state has a width of only
$\sim0.5\,\keV$~\cite{npa2004:008-010}.)

Since the quadrupole moment vanishes identically for any state with angular
momentum $J<1$, mirror pairs with ground state $J=0$ (and thus all even-even
nuclei) or $1/2$ are excluded.  This leaves us with the following mirror pairs
for consideration (circled in Fig.~\ref{fig:nuclear-chart}): $A=7$
($\isotope[7]{Li}$/$\isotope[7]{Be}$), $8$ ($\isotope[8]{Li}$/$\isotope[8]{B}$),
$9$ ($\isotope[9]{Be}$/$\isotope[9]{B}$), $9'$
($\isotope[9]{Li}$/$\isotope[9]{C}$), $11$
($\isotope[11]{B}$/$\isotope[11]{C}$), $12$
($\isotope[12]{B}$/$\isotope[12]{N}$), and $13'$
($\isotope[13]{B}$/$\isotope[13]{O}$).  Here, for odd $A$, we indicate
non-nearest neighbor mirror pairs, namely with $T_z=\pm3/2$, by placing a prime
on the value of $A$, while the rest have $T_z=\pm1/2$.  The odd-odd mirror pairs
all have the minimal $T_z=\pm1$.

Ground-state electric quadrupole moments are experimentally known (as indicated
by a ``Q'' in Fig.~\ref{fig:nuclear-chart}) for most of these nuclei, measured
variously by $\beta$ nuclear magnetic resonance, $\beta$ nuclear quadrupole
resonance, and atomic or molecular beam
measurements~\cite{stone2016:e2-moments}.  The exceptions are $\isotope[7]{Be}$,
$\isotope[9]{B}$ (the narrow resonance noted above), and $\isotope[9]{C}$.
Experimental quadrupole moments for members of the mirror pairs considered here,
as evaluated in Ref.~\cite{stone2016:e2-moments}, are summarized in
Table~\ref{tab:expt-gfmc}. Uncertainties range from the order of $10\%$ to less
than $1\%$ (although the signs of some of these quadrupole moments are
experimentally undetermined or uncertain). These experimental values provide the
basis for stringent comparisons with theoretical predictions.

The quadrupole moments of both members of the pair are thus experimentally known in
the case of the $A=8$, $11$, $12$, and $13'$ mirror pairs.  For these, the ratio
is also given for reference in Table~\ref{tab:expt-gfmc}.  For consistency in
defining the ratios, we always take the ratio of the quadrupole moment for the
proton-rich ($Z>N$) nuclide to that of the neutron-rich ($Z<N$) nuclide.

%% file: emratio_tab02.tex
\begin{table*}[t]
  \caption{Experimental ground-state quadrupole
    moments~\cite{stone2016:e2-moments} for the $p$-shell mirror pairs
    considered here, and prior \textit{ab initio} GFMC
    predictions~\cite{pastore2013:qmc-em-alt9} (discussed in
    Sec.~\ref{sec:ratio}).  Ratios within a mirror pair are deduced where
    possible, and the uncertainty on the ratio is obtained assuming the
    uncertainties on the individual moments are uncorrelated.}
  \label{tab:expt-gfmc}
\begin{center}
\begin{ruledtabular}
  \begin{tabular}{llcldddd}
    \multicolumn{4}{c}{Nuclide}&\multicolumn{2}{c}{Experiment}&\multicolumn{2}{c}{GFMC (AV18+IL7)}
    \\
    \cline{1-4}\cline{5-6}\cline{7-8}
    $A$       & & $(Z,N)$ & $J^P$ & \multicolumn{1}{c}{$Q$ ($\fm^2$)} & \multicolumn{1}{c}{Ratio} &\multicolumn{1}{c}{$Q$ ($\fm^2$)} & \multicolumn{1}{c}{Ratio} \\
    \hline
    $7$ & $\isotope[7]{Li}$ & $(3,4)$ & $3/2^-$ &
    -4.00(3)& &
    -4.0(1) &
    \\
      & $\isotope[7]{Be}$ & $(4,3)$ & $3/2^-$ &
    \multicolumn{1}{c}{---}&\multicolumn{1}{c}{---}&
    -6.7(1)\footnotemark[1]& +1.68(5)
    \\
    $8$ & $\isotope[8]{Li}$ & $(3,5)$ & $2^+$ &
    +3.14(2)& &
    +3.3(1) &
    \\
      & $\isotope[8]{B}$ & $(5,3)$ & $2^+$ &
    +6.34(14)& +2.02(5)&
    +5.9(4) & +1.79(13)
    \\
    $9$  & $\isotope[9]{Be}$ & $(4,5)$ & $3/2^-$ &
    +5.29(4)& &
    +5.1(1) &
    \\
      & $\isotope[9]{B}$ & $(5,4)$ & $3/2^-$ &
    \multicolumn{1}{c}{---}&\multicolumn{1}{c}{---}&
    +4.0(3)\footnotemark[1] & +0.78(6)
    \\
    $9'$ & $\isotope[9]{Li}$ & $(3,6)$ & $3/2^-$ &
    -3.04(2)& &
    -2.3(1) &
    \\
     & $\isotope[9]{C}$ & $(6,3)$ & $(3/2^-)$ &
    \multicolumn{1}{c}{---}&\multicolumn{1}{c}{---}&
    -4.1(4) & +1.8(2)
    \\
    $11$ & $\isotope[11]{B}$ & $(5,6)$ & $3/2^-$ &
    +4.059(10)& &
     &
    \\
       & $\isotope[11]{C}$ & $(6,5)$ & $3/2^-$ &
    \pm3.33(2)& \pm0.820(5)&
     &
    \\
    $12$ & $\isotope[12]{B}$ & $(5,7)$ & $1^+$ &
    \pm1.32(3)& &
     &
    \\
     & $\isotope[12]{N}$ & $(7,5)$ & $1^+$ &
    +1.00(9)\footnotemark[2]& \pm0.76(7)&
     &
    \\
    $13'$ & $\isotope[13]{B}$ & $(5,8)$ & $3/2^-$ &
    (+)3.65(8)& &
     &
    \\
     & $\isotope[13]{O}$ & $(8,5)$ & $3/2^-$ &
    \pm1.11(8)& \pm0.30(2) &
     &
    \\
  \end{tabular}
\end{ruledtabular}
\end{center}
\scriptsize \raggedright
\footnotemark[1] The GFMC quadrupole moments for $\isotope[7]{Be}$ and
$\isotope[9]{B}$ provided in Table~II of Ref.~\cite{pastore2013:qmc-em-alt9}
are derived under proton-neutron interchange from wave functions for the mirror
nuclides $\isotope[7]{Li}$ and $\isotope[9]{Be}$ (\textit{i.e.}, from $Q_n$,
assuming mirror symmetry).  Furthermore, from entries marked with an asterisk in
Table~II of Ref.~\cite{pastore2013:qmc-em-alt9}, indicating quadrupole moments derived
under proton-neutron interchange from the wave function for the mirror nuclide,
we may read off the calculated neutron quadrupole moments
$Q_n(\isotope[8]{Li})=+6.5(2)\,\fm^2$, $Q_n(\isotope[8]{B})=+3.0(4)\,\fm^2$,
$Q_n(\isotope[9]{Li})=-3.7(1)\,\fm^2$, and $Q_n(\isotope[9]{C})=-2.7(2)\,\fm^2$.
\\
\footnotemark[2] The quadrupole moment for $\isotope[12]{N}$ is tabulated in the
Stone~2016 evaluation~\cite{stone2016:e2-moments} as $+10.0(9)\,\fm^2$, or
$+0.100(9)\,\mathrm{b}$ in the units adopted in that tabulation. The underlying
experimental $\beta$-NMR results are those of Minamisono \textit{et
  al.}~\cite{minamisono1998:12n-quadrupole-betan-nmr}, which permit the
$\isotope[12]{N}$ quadrupole moment to be deduced relative to that of the
$\isotope[14]{N}$ reference standard.  As originally taken in conjunction with
the $\isotope[14]{N}$ quadrupole moment of Schirmacher \textit{et
  al.}~\cite{schirmacher1993:14n-hyperfine}, the experimental results yielded a
$\isotope[12]{N}$ quadrupole moment of
$+0.98(9)\,\fm^2$~\cite{minamisono1998:12n-quadrupole-betan-nmr}, as also
subsequently quoted in the Stone~2005~\cite{stone2005:m1-e2-moments} and Kelley
\textit{et al.}~2012~\cite{npa2012:011} evaluations.  However, the Stone~2016
evaluation~\cite{stone2016:e2-moments} adopts an updated $\isotope[14]{N}$
reference quadrupole moment from Pyykk\"{o}~\cite{pyykko2008:e2-moments}, which
provides a $2\%$ adjustment relative to the prior value.  This yields a
revised $\isotope[12]{N}$ quadrupole moment of $+1.00(9)\,\fm^2$, or
$+0.0100(9)\,\mathrm{b}$, as we take here.  However, the value tabulated in
Ref.~\cite{stone2016:e2-moments} differs by a shifted decimal point.
\end{table*}

%% file: emratio-ratio.tex
\section{\textit{Ab initio} predictions for quadrupole moment mirror ratios}
\label{sec:ratio}

\subsection{Mirror nuclide quadrupole moments: $A=7$}
\label{sec:ratio:a7}

\begin{figure*}
\begin{center}
\includegraphics[width=\ifproofpre{0.75}{1}\hsize]{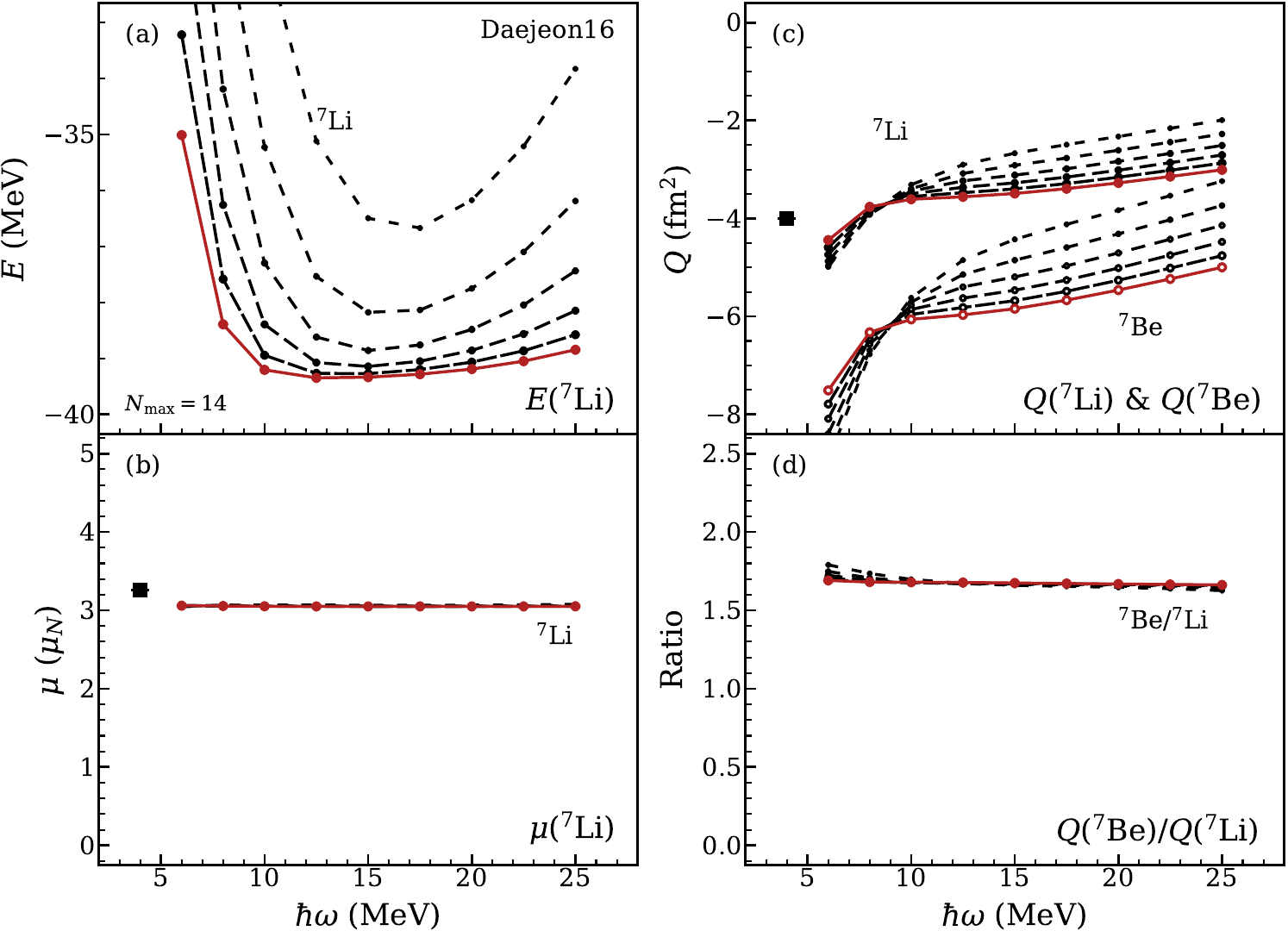}
\end{center}
\caption{Convergence of \textit{ab initio} NCCI calculations for the $A=7$
  mirror nuclides, with the Daejeon16 interaction: (a)~the $3/2^-$ ground state
  energy for $\isotope[7]{Li}$, (b)~the ground state magnetic dipole moment
  $\mu$ for $\isotope[7]{Li}$, (c)~the ground state electric quadrupole moment
  $Q$ for $\isotope[7]{Li}$ (filled symbols) and $\isotope[7]{Be}$ (open
  symbols), and (d)~the ratio of the quadrupole moment for $\isotope[7]{Be}$ to
  that of $\isotope[7]{Li}$.  Calculated values are shown as functions of the
  basis parameter $\hw$, for successive even value of $\Nmax$ (increasing symbol
  size and longer dashing), from $\Nmax=4$ (short dashed curves) to $14$ (solid curves).  
  Experimental values are also shown for the $\isotope[7]{Li}$ magnetic dipole~\cite{stone2005:m1-e2-moments} and
  electric quadrupole~\cite{stone2016:e2-moments} moments (squares, at left).
\label{fig:convergence-illustration-a7}
}
\end{figure*}
\begin{figure*}
\begin{center}
\includegraphics[width=\ifproofpre{0.75}{1}\hsize]{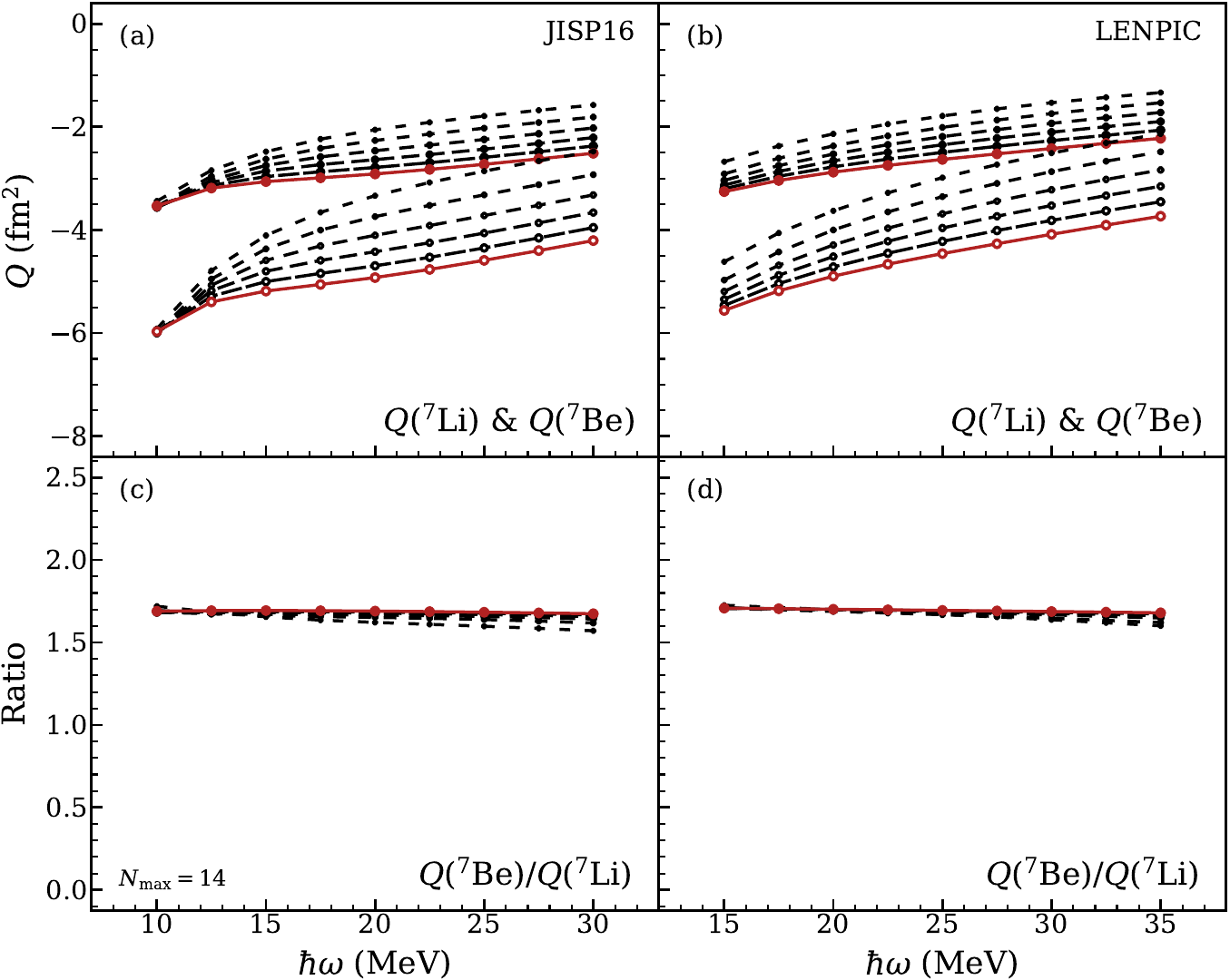}
\end{center}
\caption{Quadrupole moments for the $A=7$ mirror nuclides, calculated with the
  JISP16~(left) and LENPIC~(right) interactions: (top)~quadrupole moments for
  $\isotope[7]{Li}$ (filled symbols) and $\isotope[7]{Be}$ (open symbols), and
  (bottom)~the $\isotope[7]{Be}$ to $\isotope[7]{Li}$ mirror ratio.  Calculated
  values are shown as functions of the basis parameter $\hw$, for successive
  even values of $\Nmax$ (increasing symbol size and longer dashing), from $\Nmax=4$~(short dashed
  curves) to $14$~(solid curves).
\label{fig:qp-scan-interaction-a7}
}
\end{figure*}

Let us first consider the $A=7$ mirror pair in detail.  Here we define the NCCI
calculations, illustrate the challenges associated with predicting quadrupole
moments directly from such calculations, and then explore the rationale for instead
considering their ratios.  We will then, in subsequent subsections, survey the
results for the remaining mirror pairs, both odd-mass~(Sec.~\ref{sec:ratio:odd})
and odd-odd~(Sec.~\ref{sec:ratio:even}).

The NCCI approach~\cite{barrett2013:ncsm} is based on diagonalizing the nuclear
many-body Hamiltonian in a basis of antisymmetrized product states (Slater
determinants) constructed from some single particle basis, most commonly
harmonic oscillator orbitals.  Actual calculations must be carried out in a
finite, truncated basis.  The energies and other observables thereby obtained
are only approximations to those which would be obtained in the full many-body
space.  However, by systematically expanding the basis, it is in principle
possible to approach the full-space values to any desired degree of accuracy.
The actual accuracy which can be reached is subject to computational limitations
on the problem size.

The NCCI basis states may be organized according to the number $\Nex$ of
oscillator excitations relative to the lowest Pauli-allowed filling of
oscillator shells.  The many-body basis is then, for practical reasons (\textit{e.g.},
Ref.~\cite{caprio2020:intrinsic}), conventionally
truncated to a maximum number $\Nmax$ of oscillator excitations above the lowest
filling (\textit{i.e.}, to $\Nex\leq\Nmax$).  The space spanned by the many-body
basis also depends upon the oscillator length $b$ of the underlying oscillator
orbitals, commonly expressed in terms of the corresponding oscillator energy
$\hw$~\cite{suhonen2007:nucleons-nucleus}.  The full-space
values for observables should be recovered as $\Nmax$ approaches infinity, in
which limit the calculated values also become independent of $\hw$.

Consider first the calculated energy eigenvalue for the ground state of
$\isotope[7]{Li}$, shown in Fig.~\ref{fig:convergence-illustration-a7}(a), for
the Daejeon16 interaction.  (The present calculations are obtained using the $M$-scheme NCCI code
MFDn~\cite{aktulga2013:mfdn-scalability,shao2018:ncci-preconditioned}.)
Each curve represents calculations at fixed
truncation $\Nmax$ ($\Nmax=4$ to~$14$, even), for varying $\hw$.  The calculated
ground state energy is variationally bounded below by the full-space value.  An
approach to convergence is evident in the curves becoming successively closer
together (independent of $\Nmax$) and flatter (independent of $\hw$), as they
approach the variational lower bound, such that the highest $\Nmax$ curves lie
nearly atop each other at the scale shown.  A qualitatively similar convergence
pattern is obtained for other choices of interaction, although the rate of
convergence and the location (in $\hw$) of the variational minimum in each curve
will in general differ.

The quality of convergence in an NCCI calculation depends upon the observable
(and states) under consideration.  For instance, magnetic dipole ($M1$)
observables, which are largely sensitive to the angular momentum structure of
the wave functions, may attain very rapid convergence, as illustrated for the
calculated dipole moment of $\isotope[7]{Li}$ in
Fig.~\ref{fig:convergence-illustration-a7}(b).  (The calculated magnetic dipole
moment is also reasonably consistent with the experimental
value~\cite{stone2005:m1-e2-moments}.)  In contrast, electric quadrupole ($E2$)
observables, which are more sensitive to the radial behavior (the ``tails'' of
the wave functions are amplified by the $r^2$ radial dependence of the
quadrupole operator), are comparatively difficult to converge.

Indeed, the calculated quadrupole moment for the $\isotope[7]{Li}$ ground state,
shown by the filled symbols in Fig.~\ref{fig:convergence-illustration-a7}(c),
depends strongly upon $\hw$ at fixed $\Nmax$, especially for low $\Nmax$, and
continues to vary steadily with increasing $\Nmax$.  Note that the overall trend
of each curve with respect to $\hw$, from asymptotically large magnitude as
$\hw\rightarrow 0$ to asymptotically small magnitude as $\hw\rightarrow\infty$,
is simply a consequence of the dependence on $\hw$ of the length scale of the
basis functions.  For a single oscillator basis function, the quadrupole moment
scales as $Q\propto b^2$, where $b\propto (\hw)^{-1/2}$, and is thus inversely
proportional to $\hw$.

Encouragingly, there does appear to be some progress towards convergence, with
increasing $\Nmax$, in the calculated $\isotope[7]{Li}$ quadrupole moment.  In
particular, note the significant flattening (or ``shouldering'') of the curves
at high $\Nmax$, as well as some degree of compression of successive curves
against each other with increasing $\Nmax$.  There is also a narrow region,
around $\hw\approx9\,\MeV$, where the curves cross each other, and therefore
locally approach $\Nmax$ independence.  Such crossings have been suggested as a
heuristic indicator of
convergence~\cite{nogga2006:7li-ncsm-chiral,bogner2008:ncsm-converg-2N,cockrell2012:li-ncfc}
(although, in general, such crossings drift with
$\Nmax$~\cite{caprio2014:cshalo} and are thus of limited use in estimating the
converged value).  It is perhaps reassuring that, in the shoulder region, the
calculated values of the quadrupole moment appear to be roughly consistent with,
and approaching, the experimental value.
Nonetheless, it is less than obvious
how to extract a firm quantitative value for the predicted quadrupole moment in
the full, untruncated many-body problem (at least without significant further
assumptions about the nature of the convergence, \textit{e.g.}, as in
Ref.~\cite{odell2016:ir-extrap-quadrupole}).

The calculated quadrupole moment for the mirror nuclide $\isotope[7]{Be}$, shown
by the open symbols in Fig.~\ref{fig:convergence-illustration-a7}(c), shares a
similarly encouraging approach to convergence but is likewise insufficiently
converged to provide a firm quantitative prediction.  Indeed, the two
quadrupole moments in Fig.~\ref{fig:convergence-illustration-a7}(c) have similar
convergence trends, differing primarily in overall normalization.

Herein lies the motivation for considering ratios.  The error introduced by
basis truncation may largely cancel in the ratio of these quadrupole moments,
providing a more robustly converged prediction for their ratio.  Such
cancellation and robust convergence has already been noted for the ratio between
the mirror $E2$ transitions ($3/2^-\rightarrow 1/2^-$) in these two
nuclei~\cite{henderson2019:7be-coulex} (see Fig.~6 of
Ref.~\cite{henderson2019:7be-coulex}).  Indeed the ratio of the calculated
quadrupole moments for $\isotope[7]{Be}$ and $\isotope[7]{Li}$, shown in
Fig.~\ref{fig:convergence-illustration-a7}(d), is seen to be largely
independent of $\Nmax$ and $\hw$.

It is informative to compare the convergence behavior obtained for different
choices of interaction, derived by different procedures.  The Daejeon16
interaction~\cite{shirokov2016:nn-daejeon16} is a comparatively ``soft''
interaction.  It is based on the two-body part of the Entem-Machleidt (EM)
\nthreelo{} chiral EFT interaction~\cite{entem2003:chiral-nn-potl}, softened via
a similarity renormalization group (SRG)
transformation~\cite{bogner2007:srg-nucleon} to enhance convergence, and then
adjusted via a phase-shift equivalent transformation to better describe light
nuclei with $A\leq16$ (see Ref.~\cite{maris2019:daejeon16-lenpic-pshell-ntse18}
for comparison with experiment).  The earlier JISP16
interaction~\cite{shirokov2007:nn-jisp16} is derived instead from
nucleon-nucleon scattering data by $J$-matrix inverse scattering, then similarly
adjusted via phase-shift equivalent transformations to data for nuclei with
$A\leq16$ (see Ref.~\cite{maris2013:ncsm-pshell} for comparison with
experiment).  Alternatively, the LENPIC
interaction~\cite{epelbaum2015:lenpic-n4lo-scs,epelbaum2015:lenpic-n3lo-scs} is
a newer chiral EFT interaction, developed with an ultraviolet regularization
scheme designed to minimize finite-cutoff artifacts.  We consider the two-body
part of this interaction at \ntwolo{}, which provides a reasonable description
of nuclear observables without
adjustment~\cite{binder2016:lenpic-chiral,binder2018:lenpic-scs-light}.  Here we
take it with a semi-local coordinate-space regulator ($R=1\,\fm$), and in its
``bare'' form, \textit{i.e.}, without subsequent SRG transformation.

The calculated $A=7$ quadrupole moments, and their ratio, for the JISP16 and
LENPIC interactions are shown in Fig.~\ref{fig:qp-scan-interaction-a7}.  (For
each interaction, we consider calculations in an $\hw$ range centered on the
approximate location of the variational minimum of the ground state energy, for
calculations with that interaction.) The calculated quadrupole
moments themselves show less indication of convergence than for Daejeon16
[Fig.~\ref{fig:convergence-illustration-a7}(c)].  There is a hint of shouldering
in the calculated quadrupole moments for JISP16
[Fig.~\ref{fig:qp-scan-interaction-a7}(a)], which cross at $\hw\lesssim
10\,\MeV$, but little suggestion of convergence at all for the unsoftened LENPIC
interaction [Fig.~\ref{fig:qp-scan-interaction-a7}(b)].

Nonetheless, the mirror
ratio of quadrupole moments [Fig.~\ref{fig:qp-scan-interaction-a7}(c,d)] is
again robustly converged, much as for the Daejeon16 interaction
[Fig.~\ref{fig:convergence-illustration-a7}(d)]. We observe close quantitative
agreement between the predictions for the $A=7$ quadrupole moment ratio obtained
with the Daejeon16, JISP16, and LENPIC interactions.

Within the $A=7$ mirror pair, the quadrupole moment of $\isotope[7]{Be}$ is unmeasured, and thus
experiment does not provide a test of the NCCI predictions.  Rather, the robust
NCCI prediction of the ratio (approximately $+1.7$), taken in conjunction with the measured
quadrupole moment of $-4.0\,\fm^2$ for $\isotope[7]{Li}$ (Table~\ref{tab:expt-gfmc}), provides a
concrete prediction of approximately $-6.8\,\fm^2$ for the unmeasured $\isotope[7]{Be}$ quadrupole
moment.

\subsection{Odd-mass mirror pairs}
\label{sec:ratio:odd}

\begin{figure*}
\begin{center}
\includegraphics[width=\ifproofpre{1}{0.95}\hsize]{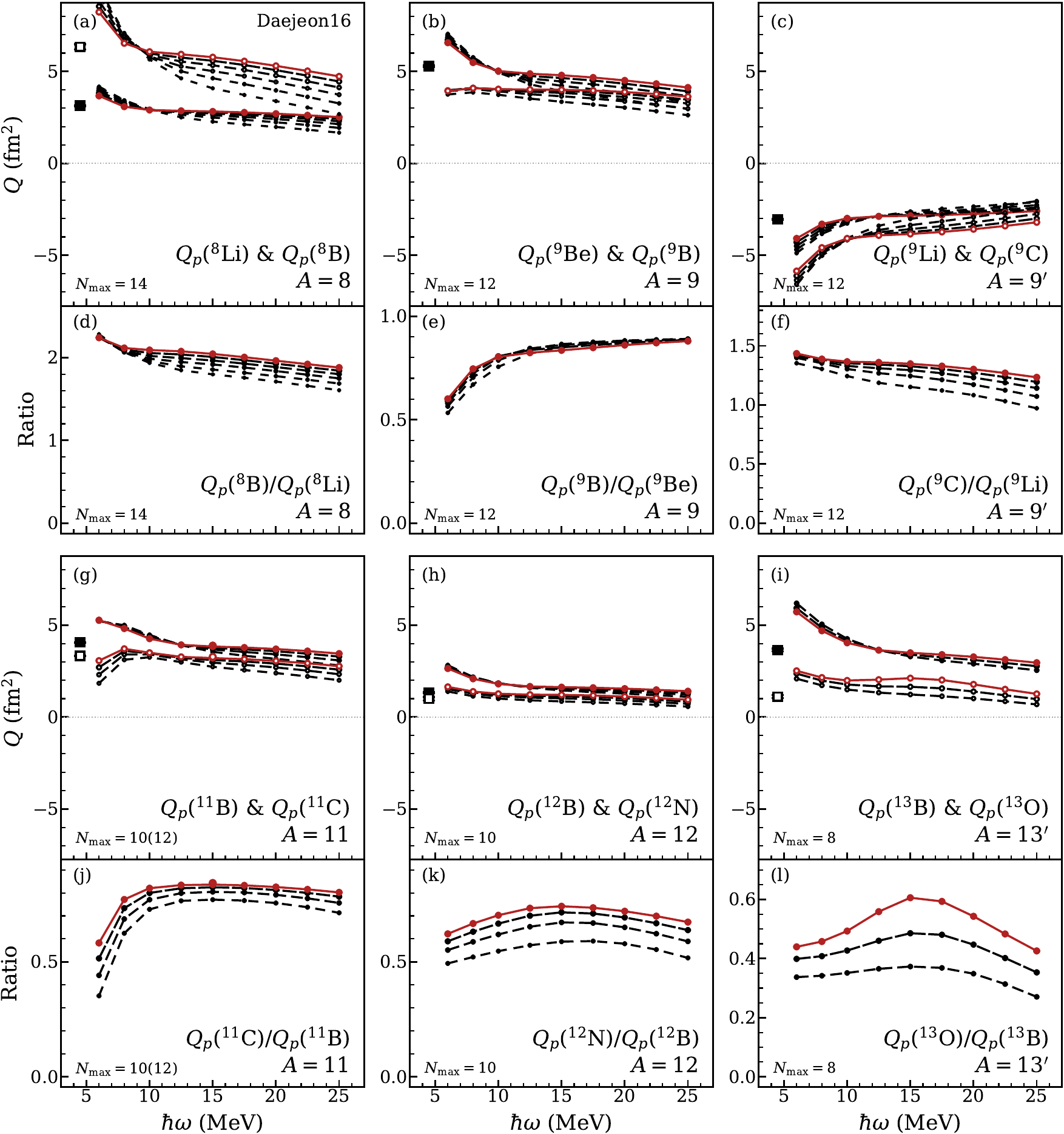}
\end{center}
\caption{Quadrupole moments for $p$-shell mirror nuclides (upper member of each
  pair of panels), for $A>7$, and the ratio of quadrupole moments across each
  mirror pair (lower member of each pair of panels), calculated with the
  Daejeon16 interaction.  Calculated values (circles) are shown as functions of
  the basis parameter $\hw$, for successive even values of $\Nmax$ (increasing
  symbol size and longer dashing), from $\Nmax=4$~(short dashed curves) to the maximum value for that mirror pair~(solid
  curves), indicated at bottom.  Quadrupole moments for the neutron-rich ($Z<N$) nuclides are shown
  with filled symbols, and those for the proton-rich ($Z>N$) nuclides with open
  symbols.  Also shown are the experimental quadrupole
  moments~\cite{stone2016:e2-moments} (squares) from Table~\ref{tab:expt-gfmc}
  (some signs experimentally undetermined).
  \label{fig:mirror-scan-agt7}
}
\end{figure*}

Let us now consider the \textit{ab initio} predictions for the quadrupole
moments and their ratios for the remaining $p$-shell mirror pairs ($A>7$).  The
calculated quadrupole moments for both members of each mirror pair are overlaid
in the upper member of each pair of panels in Fig.~\ref{fig:mirror-scan-agt7},
demonstrating the convergence behavior with respect to $\Nmax$ and $\hw$.  Then
the calculated ratio is shown in the lower member of each pair of panels in
Fig.~\ref{fig:mirror-scan-agt7}.  We again focus on the calculations with the
Daejeon16 interaction for detailed analysis (comprehensive tabulations of
the calculated quadrupole moments for all three interactions are provided in the
Supplemental Material~\cite{supplemental-material}).  The experimental values are
indicated for comparison, where available, in Fig.~\ref{fig:mirror-scan-agt7},
with the caveat that signs are not always experimentally determined (see
Table~\ref{tab:expt-gfmc}).  While calculations for all of the mirror pairs
are shown in Fig.~\ref{fig:mirror-scan-agt7}, we discuss first the remaining
odd-$A$ nuclides, before turning to the odd-odd nuclides below in
Sec.~\ref{sec:ratio:even}.

\begin{figure*}
\begin{center}
\includegraphics[width=\ifproofpre{0.75}{1}\hsize]{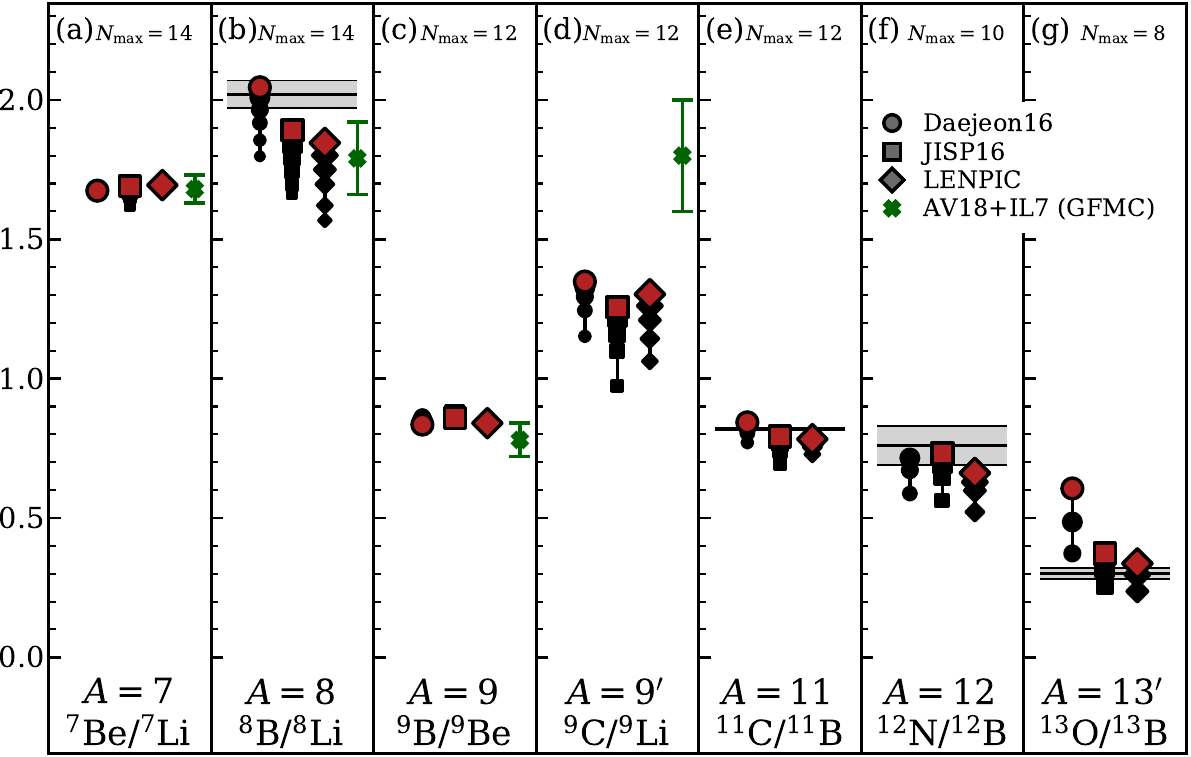}
\end{center}
\caption{Quadrupole moment mirror ratios for $p$-shell mirror pairs, obtained
  with the Daejeon16~(circles), JISP16~(squares), and LENPIC~(diamonds)
  interactions, at fixed $\hw$ ($15\,\MeV$, $20\,\MeV$, and $25\,\MeV$,
  respectively, for the three interactions).  Calculated values are shown for
  successive even values of $\Nmax$ (increasing symbol size), from $\Nmax=4$ to
  the maximum value for that mirror pair, indicated at top.  Also shown (from
  Table~\ref{tab:expt-gfmc}) are the GFMC AV18+IL7
  predictions~\cite{pastore2013:qmc-em-alt9} (crosses) and ratios of
  experimental quadrupole moments~\cite{stone2016:e2-moments} (horizontal line
  and error band), where some signs are experimentally undetermined.
\label{fig:mirror-ratio-teardrop}
}
\end{figure*}

To provide a concise, though less nuanced, global comparison of results for the
quadrupole moment ratios, for all mirror pairs and for all choices of
interaction, it is helpful to take a slice at fixed $\hw$ through the
convergence results, considering only the $\Nmax$ dependence.  A natural choice
for $\hw$ is again the approximate location of the variational minimum of the
ground state energy [recall the $\isotope[7]{Li}$ energy calculations in
  Fig.~\ref{fig:convergence-illustration-a7}(a)].  The exact position of this
minimum varies with nuclide and with $\Nmax$, but for Daejeon16 we standardize
on $\hw=15\,\MeV$ as a reasonable estimate, and similarly we adopt $20\,\MeV$
for JISP16 and $25\,\MeV$ for LENPIC.  Then we are able to make side-by-side
numerical comparisons of the $\Nmax$-dependence of the calculated quadrupole
moment ratios, as shown in Fig.~\ref{fig:mirror-ratio-teardrop}.  The calculated
ratio for the $A=7$ mirror pair, discussed above in Sec.~\ref{sec:ratio:a7}, may
be found at far left [Fig.~\ref{fig:mirror-ratio-teardrop}(a)].

Furthermore, for the lighter nuclides ($A\leq9$), \textit{ab initio}
results~\cite{pastore2013:qmc-em-alt9} for the quadrupole moments have
previously been obtained by Green's function Monte Carlo (GFMC)
methods~\cite{carlson2015:qmc-nuclear}, in calculations based on the Argonne
$v_{18}$ (AV18) two-nucleon~\cite{wiringa1995:nn-av18} and Illinois-7 (IL7)
three-nucleon~\cite{pieper2008:3n-il7-fm50} potentials.  The predicted quadrupole
moments are quoted for reference in Table~\ref{tab:expt-gfmc}.  For the $A=7$
pair, the resulting ratio, shown as a cross in
Fig.~\ref{fig:mirror-ratio-teardrop}(a), is closely aligned with the present
predictions, to within the GFMC statistical uncertainties.

In the $A=9$ mirror pair ($\isotope[9]{Be}$ and $\isotope[9]{B}$), the
calculated quadrupole moments themselves
[Fig.~\ref{fig:mirror-scan-agt7}(b)] show a trend towards convergence
with increasing $\Nmax$, much as for the $A=7$ pair above
[Fig.~\ref{fig:convergence-illustration-a7}(c)] (when comparing, note the
opposite overall sign of the quadrupole moments, between $A=7$ and $A=9$).  The
``shoulder'' in the calculated quadrupole moment for $\isotope[9]{Be}$ (filled
symbols) suggests a value roughly consistent with experiment.  For
$\isotope[9]{B}$ (open symbols), the turnover in the calculated quadrupole
moment for $\hw \lesssim10\,\MeV$ appears to reflect a more general breakdown in
the ability of the calculation to preserve the excitation spectrum of $3/2^-$
states at low $\hw$ (bases with excessively low $\hw$ have an excessively long
oscillator length scale).

While these individual calculated quadrupole moments still exhibit significant
basis dependence, the calculated ratio
[Fig.~\ref{fig:mirror-scan-agt7}(e)] is largely independent of $\hw$
and $\Nmax$, even at comparatively low $\Nmax$, for $\hw\gtrsim
10\,\MeV$.  (The failure below $\hw\approx 10\,\MeV$ is perhaps not surprising,
given the more general breakdown noted above.) From
Fig.~\ref{fig:mirror-ratio-teardrop}(c), it is seen that the Daejeon16, JISP16, and
LENPIC interactions yield consistent predictions for the quadrupole moment
ratio, furthermore consistent with the GFMC AV18+IL7 results.  The stable and
interaction-independent NCCI prediction of the ratio (in the range of
approximately $+0.80$ to $+0.85$), taken in conjunction with the measured
quadrupole moment of $+5.3\,\fm^2$ for $\isotope[9]{Be}$
(Table~\ref{tab:expt-gfmc}), provides a prediction of about $+4.2\,\fm^2$ to
$+4.5\,\fm^2$ for the unmeasured $\isotope[9]{B}$ quadrupole moment.

In the $A=9'$ mirror pair ($\isotope[9]{Li}$ and $\isotope[9]{C}$), the
convergence of the quadrupole moment ratio is less robust.  The
moments themselves [Fig.~\ref{fig:mirror-scan-agt7}(c)] again show
signs of shouldering, and for $\isotope[9]{Li}$ (filled symbols) there is again
general consistency with the experimental value.  While the quadrupole moment
ratio [Fig.~\ref{fig:mirror-scan-agt7}(f)] would also appear to be
approaching convergence, this occurs more slowly than in the previous
cases. Consequently, one would be hard put to make an estimate except that the
value appears likely to be between $1.3$ and $1.5$ with Daejeon16.  The different interactions
[Fig.~\ref{fig:mirror-ratio-teardrop}(d)] yield a spread of $\sim 10\%$ in the
predicted ratio, in the calculations at the highest $\Nmax$.  However, this
spread also seems to be decreasing with increasing $\Nmax$, so it might not
reflect a true difference in predictions for the different interactions.

The direction of convergence of the NCCI predicted ratios for the $A=9'$ mirror
pair is such as to bring them towards consistency with the GFMC AV18+IL7 result.
Note the larger statistical uncertainties obtained in the GFMC calculations for
the $A=9'$ mirror nuclides, in this case accompanying the poorer convergence of
the NCCI calculations.

Note that $\isotope[9]{C}$ is a comparatively weakly-bound nucleus, with its
ground state only $\sim1.3\,\MeV$ below the $\isotope[8]{B}+p$ breakup
threshold~\cite{npa2004:008-010}, relative to its mirror nuclide
$\isotope[9]{Li}$, which is bound by $\sim4.1\,\MeV$ against the analogous
$\isotope[8]{Li}+n$ breakup channel.  This difference could ostensibly introduce
greater sensitivities in calculating the long-range behavior of the wave
function, \textit{e.g.}, through an extended proton single-particle wave
function.  Yet, binding energy alone clearly does not suffice to explain the
differences, as $\isotope[9]{B}$ is $\sim0.3\,\MeV$ \textit{above} the
$\isotope[8]{Be}+p$ threshold, and comparatively rapid convergence was obtained
above.  (For that matter, in the $A=7$ pair considered above, $\isotope[7]{Be}$
is bound by not much more, $\sim1.6\,\MeV$~\cite{npa2002:005-007}, although this
is against a cluster breakup mode, $\isotope[3]{He}+\isotope[4]{He}$, rather
than separation of a single nucleon.)

The $A=11$ pair ($\isotope[11]{C}$ and $\isotope[11]{B}$) presents a
comparatively simple picture of convergence.  For both members of the pair, the
quadrupole moments themselves [Fig.~\ref{fig:mirror-scan-agt7}(g)] both show
some evidence of shouldering, consistent with the experimental value.  In the
calculations for Fig.~\ref{fig:mirror-scan-agt7}(g), full scans over $\hw$ are
carried out through $\Nmax=10$.  However, an $\Nmax=12$ calculation, currently
just computationally feasible, is included at the variational minimum $\hw$
[hence the indication  of $\Nmax$ as ``$10(12)$'' in Fig.~\ref{fig:mirror-scan-agt7}(g,j)
  and subsequent figures].  The now-familiar turnover at low $\hw$
($\lesssim10\,\MeV$), is observed for the $\isotope[11]{C}$ quadrupole moment.
These nuclei, incidentally, are the most bound of the mirror pairs considered
here, by over $11\,\MeV$ for $\isotope[11]{B}$ and over $7\,\MeV$ for
$\isotope[11]{C}$~\cite{npa2012:011}.

The ratio of quadrupole moments for $A=11$
[Fig.~\ref{fig:mirror-scan-agt7}(j)] is rapidly converging, approaching
a value in the vicinity of $0.8$.  It appears that a slightly higher value may
be obtained with the Daejeon16 interaction than with the others
[Fig.~\ref{fig:mirror-ratio-teardrop}(d)], but all lie within a range of $0.7$ to
$0.9$, closely consistent with experiment.

Finally, among the odd-mass nuclei, the $A=13'$ pair ($\isotope[13]{O}$ and
$\isotope[13]{B}$) presents a qualitatively different convergence behavior.
Examining the calculated quadrupole moments
[Fig.~\ref{fig:mirror-scan-agt7}(i)], we see that those for
$\isotope[13]{B}$ (filled circles) exhibit the familiar shouldering and again
tend toward a value consistent with experiment.  However, the corresponding
curves for $\isotope[13]{O}$ (open circles) bulge upwards with increasing
$\Nmax$.  At $\hw=15\,\MeV$, the value approximately doubles between $\Nmax=4$
and $\Nmax=8$.  These calculated values already exceed the experimental
quadrupole moment and continue to move away with increasing $\Nmax$.  The quadrupole moment ratio
[Fig.~\ref{fig:mirror-scan-agt7}(l)] likewise rapidly grows past the
experimental ratio [Fig.~\ref{fig:mirror-ratio-teardrop}(g)].  (The ratios obtained
with the JISP16 and LENPIC interactions similarly grow steadily, though not as
rapidly, with increasing $\Nmax$~\cite{supplemental-material}.)

In a shell-model picture, the $A=13'$ nuclei are semi-magic.  The protons in
$\isotope[13]{O}$ form a closed major oscillator shell in a $0\hw$
configuration.  Thus, the proton quadrupole moment in $\isotope[13]{O}$ must
vanish at $\Nmax=0$.  (The neutron quadrupole moment in $\isotope[13]{B}$ must
similarly vanish at $\Nmax=0$.)  Any nonzero value of the ratio of the
$\isotope[13]{O}$ quadrupole moment to the $\isotope[13]{B}$ quadrupole moment
must come from the introduction of $2\hw$ or higher configurations into the NCCI
wave function, as $\Nmax$ increases from $0$.  Thus, the semimagic nature of
$\isotope[13]{O}$ at least explains why the convergence behavior for the proton
quadrupole moment [open symbols in Fig.~\ref{fig:mirror-scan-agt7}(i)] must
qualitatively differ from the other cases considered.

However, semimagicity does not immediately explain why the calculated quadrupole
moment continues to grow towards excessively large values with increasing
$\Nmax$.  Experimentally, the ground state of $\isotope[13]{O}$ is only
$\sim1.5\,\MeV$ below the breakup threshold
($\isotope[12]{N}+p$)~\cite{npa1991:013-015}, so one might suspect sensitivity
to long-range behavior of the wave function, but the same convergence pattern is
calculated for the neutron quadrupole moment in $\isotope[13]{B}$, which is more
tightly bound, by $\sim4.9\,\MeV$ with respect to $\isotope[12]{B}+n$.  An
alternative explanation in terms of shape coexistence is noted below
(Sec.~\ref{sec:discussion}).

\subsection{Odd-odd mirror pairs}
\label{sec:ratio:even}

Returning now to the odd-odd nuclides, let us consider the $A=8$ pair
($\isotope[8]{Li}$ and $\isotope[8]{B}$).  Although both calculated quadrupole
moments [Fig.~\ref{fig:mirror-scan-agt7}(a)] ultimately display the familiar
shouldering behavior for high $\Nmax$, at values roughly consistent with
experiment, the convergence for $\isotope[8]{B}$ (open circles) is notably
slower than for most of the others in Fig.~\ref{fig:mirror-scan-agt7} (note the
wider spread in calculated curves).  Such slow convergence is perhaps not
surprising given the expected proton halo structure of
$\isotope[8]{B}$~\cite{minamisono1992:8b-quadrupole-beta-nmr,kitagawa1993:shell-8b-17f-quadrupole-moment-halo,henninger2015:8b-fmd},
and thus extended tail region to the proton distribution.  Here it is worth
noting the $\isotope[8]{B}$ ground state is extremely weakly bound, compared to
any of the other (bound) nuclei considered here, by only $\sim0.13\,\MeV$ with
respect to $\isotope[7]{Be}+p$ breakup~\cite{npa2004:008-010}, while
$\isotope[8]{Li}$, the neutron-rich member of this mirror pair, is bound by
$\sim2.0\,\MeV$ with respect to $\isotope[7]{Li}+n$ breakup.

The ratio of quadrupole moments for $A=8$ [Fig.~\ref{fig:mirror-scan-agt7}(d)]
is therefore also slow to converge.  Although the curves at high $\Nmax$ form
shoulders, or plateaus, these plateaus continue to move slowly but steadily
upward with increasing $\Nmax$.  The bulk of this $\Nmax$ dependence comes from
the $\isotope[8]{B}$ quadrupole moment.  

The resulting ratios obtained with all three interactions
[Fig.~\ref{fig:mirror-ratio-teardrop}(b)] are in the vicinity of the
experimental ratio. The Daejeon16 result for the ratio is just growing past the
experimental value, at the highest $\Nmax$ calculated, while the JISP16 and
LENPIC results are still approaching it.  The continued basis sensitivity
precludes more precise comparison.  Here again, as for the $A=9'$ pair above
(Sec.~\ref{sec:ratio:odd}), the slow convergence of the ratio in the NCCI
calculations is accompanied by a large GFMC statistical uncertainty.

For the $A=12$ pair ($\isotope[12]{B}$ and $\isotope[12]{N}$), the experimental quadrupole
moments are markedly smaller in magnitude (by a factor of $\sim3$--$6$) than
those considered above, except for the semi-magic $\isotope[13]{O}$.
Consider, then, the calculated quadrupole moments
[Fig.~\ref{fig:mirror-scan-agt7}(h)].  Those for $\isotope[12]{B}$
(filled circles) show shouldering, and would appear to be converging, but tend
towards a value perhaps as much as $50\%$ higher than the experimental value.
Those for $\isotope[12]{N}$ (open circles), while also showing some
sort of shouldering, only gradually come closer together with increasing
$\Nmax$, showing no clear sign of convergence, but have also already grown
past the experimental value.

Given the different convergence patterns for the individual quadrupole moments,
the calculated ratio [Fig.~\ref{fig:mirror-scan-agt7}(k)] for the
$A=12$ pair is consequently also comparatively slow to converge, but the
calculated values do appear to form a roughly geometric progression, approaching
a ratio $\sim0.8$.  There is a spread of $\sim10\%$ among the calculated ratios
obtained with the three different interactions
[Fig.~\ref{fig:mirror-ratio-teardrop}(f)], at the highest $\Nmax$, but all would
seem on course to yield values within the uncertainties of the experimental
ratio.

The $\isotope[12]{N}$ ground state is bound by only $\sim0.60\,\MeV$ with
respect to $\isotope[11]{C}+p$ breakup~\cite{npa1990:011-012}.  One might be
tempted to attribute slow convergence for the $A=12$ quadrupole moment ratio,
and likewise for the $A=8$ ratio above, to this weak binding of the proton-rich
member of the pair, which may well contribute.  However, odd-odd nuclei have
notoriously complicated spectra~\cite{casten2000:ns} even when strongly bound,
and a comparatively high level density near the ground state is conducive to
mixing.  Arguments to this effect for $\isotope[8]{Li}$, in particular, may be
made~\cite{hendersonxxxx:8li-coulex} on the basis of the appearance of
approximately degenerate ground states in an Elliott
$\grpsu{3}$~\cite{elliott1958:su3-part1,*elliott1958:su3-part2,*elliott1963:su3-part3,*elliott1968:su3-part4,harvey1968:su3-shell}
symmetry description.

%% file: emratio-symm.tex
\section{Mirror symmetry breaking}
\label{sec:symm}

\begin{figure*}
  \begin{center}
    \includegraphics[width=\ifproofpre{1}{1}\hsize]{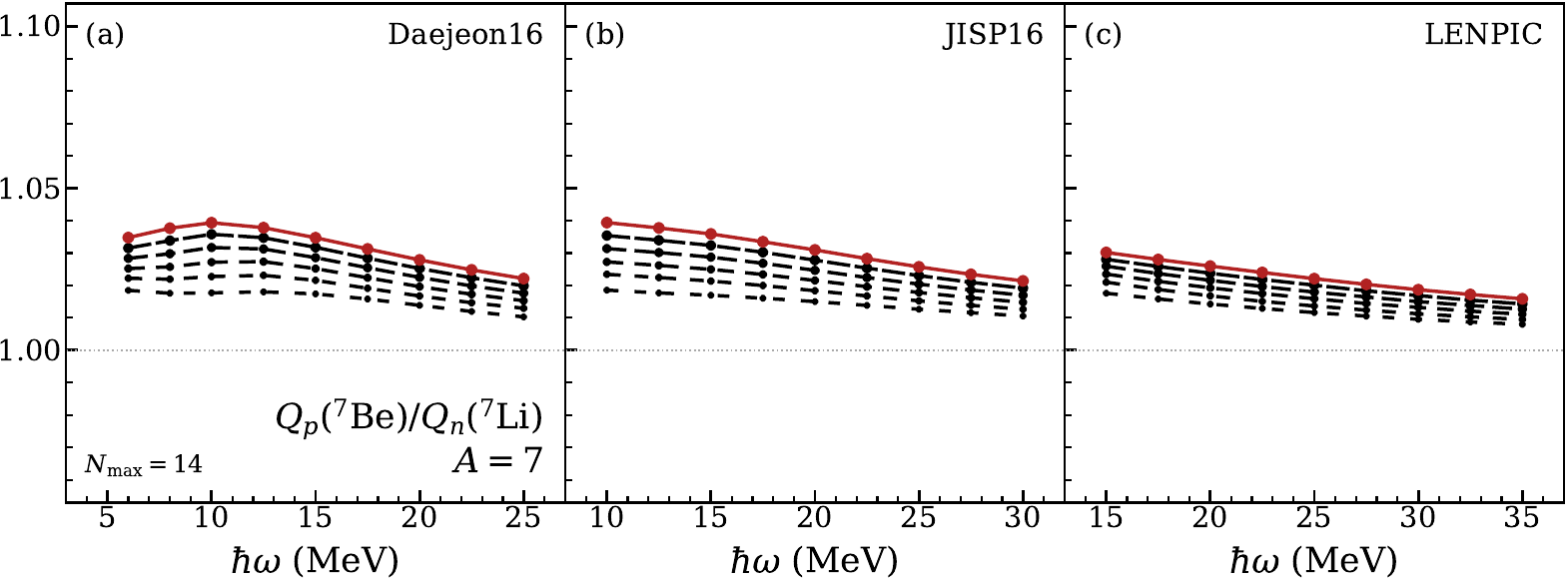}
    \\[1ex]
    \includegraphics[width=\ifproofpre{1}{1}\hsize]{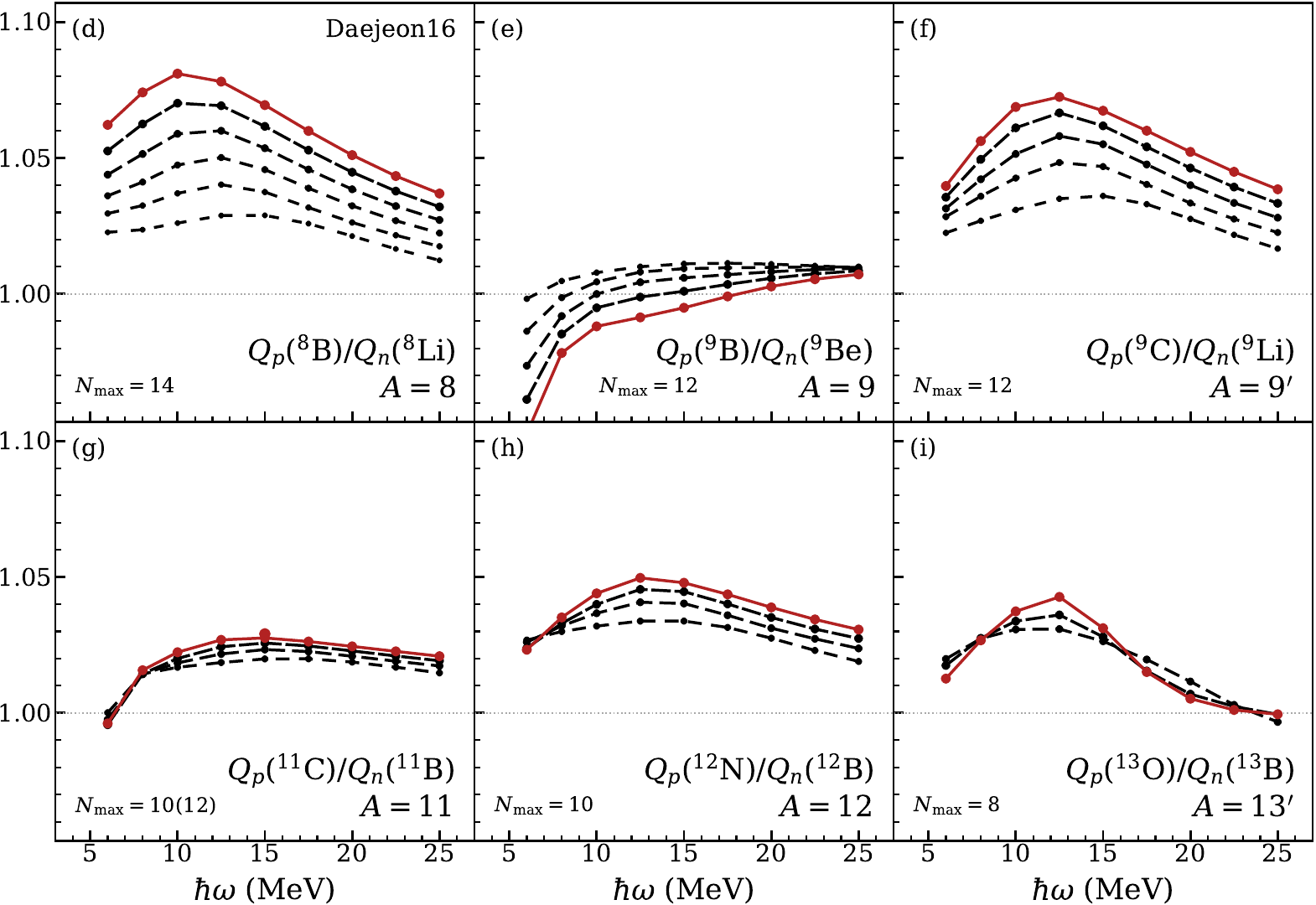}
  \end{center}
\caption{Relation between $Q_n$ (in the proton-rich nuclide) to its mirror proxy $Q_p$ (in the neutron-rich nuclide), represented as the ratio $\tilde{Q}_p/Q_n$ (see text):
  (top)~for the $A=7$ mirror pair with Daejeon16, JISP16, and LENPIC
  interactions and (bottom)~for the $A>7$ mirror pairs with the Daejeon16
  interaction.  Calculated values (circles) are shown as functions of
  the basis parameter $\hw$, for successive even values of $\Nmax$ (increasing
  symbol size and longer dashing), from $\Nmax=4$~(short dashed curves) to the maximum value for that mirror pair~(solid
  curves), indicated at bottom.
\label{fig:mirror-asymmetry-ratio-scan}
}
\end{figure*}

Let us now return to the question of the extent to which proton and neutron
quadrupole moments are actually related by mirror symmetry across a mirror pair.
In the present \textit{ab initio} calculations, the isospin is found to be a
``good'' quantum number, with isospin $T=\abs{T_z}$, to high precision, for all the
calculated ground state wave functions.  Wave function components with
$T>\abs{T_z}$ contribute, \textit{e.g.}, $\lesssim 10^{-4}$ to the norm of the calculated ground state wave function for $\isotope[7]{Li}$.\footnote{The isospin
  contamination is most simply ascertained by evaluating the expectation value
  of $\vec{T}^2$ within the many-body wave function.  For a state which is an
  admixture of $T=\abs{T_z}$ and $T=\abs{T_z}+1$ components,
  \begin{math}
    \tket{\Psi}=\alpha\tket{\Psi_{T=\abs{T_z}}}+\beta\tket{\Psi_{T=\abs{T_z}+1}},
  \end{math}
with
  probabilities $\alpha^2$ and $\beta^2$, respectively, it is readily verified
  that the $T=\abs{T_z}+1$ component contributes with probability
\begin{math}
  \beta^2=[2(\abs{T_z}+1)]^{-1}[\tbracket{\vec{T}^2}-\abs{T_z}(\abs{T_z}+1)].
\end{math}
More generally, if components with multiple $T\geq\abs{T_z}$ contribute, this two-component
estimate places an upper bound on the total $T> \abs{T_z}$ contribution.}

While the observation that each ground state in the mirror pair has good isospin
is sufficient to exclude isospin \textit{mixing}, it does not in itself
establish isospin (or mirror) \textit{symmetry} between the states.  Under
isospin symmetry, an isobaric multiplet consists of states which not only share
the same $T$, but also, more specifically, have wave functions related by
successive application of the isospin raising and lowering operators.  (For
mirror states, the wave functions are, equivalently, related by a rotation by
$\pi$ in isospin space.)  Even if the isospin violating part of the many-body
Hamiltonian does not significantly induce mixing of different isospins in the
wave function, it might still induce mixing within the $T=\abs{T_z}$ sector of
the many-body space in a way which varies with $T_z$, and thus differs across
the analog states.  This would be apparent in a violation of the isospin (or
mirror) symmetry predictions for observables.  For mirror quadrupole moments in
particular, there may be deviations from equality between the neutron quadrupole
moment $Q_n$ of one mirror state and the proton quadrupole moment $Q_p$ of the
other.

We have thus far considered the ratio of the proton (electric) quadrupole
moments across the mirror pair (Sec.~\ref{sec:ratio}), partly as an experimental
observable in its own right to provide a test for \textit{ab initio} theory, but
partly also on the premise it may be taken as a proxy for the ratio $Q_n/Q_p$
within a single nuclide, to directly reflect proton-neutron structure in that
nuclide.  If we denote the quadrupole moment in the ``mirror'' nuclide by
$\tilde{Q}_p$, our specific interest is thus in how well the mirror ratio
$\tilde{Q}_p/Q_p$ approximates $Q_n/Q_p$. These ratios are related by
\begin{equation}
  \label{eqn:ratio-relation}
  \frac{\tilde{Q}_p}{Q_p}
  =
  \biggl(
  \frac{\tilde{Q}_p}{Q_n}
  \biggr)\biggl(
  \frac{Q_n}{Q_p}
  \biggr)
  ,
\end{equation}
where the deviation of the first factor, $\tilde{Q}_p/Q_n$, from unity thus measures how much error
is introduced by taking the mirror ratio as a proxy for the neutron-proton
ratio.  Based on the present wave functions from NCCI calculations, we therefore
explicitly calculate $\tilde{Q}_p/Q_n$ and look for deviations from unity.

Let us start with the $A=7$ mirror pair, and explore the dependence of the
calculated $\tilde{Q}_p/Q_n$~--- in this case,
$Q_p(\isotope[7]{Be})/Q_n(\isotope[7]{Li})$~--- on the basis parameters of the
truncated calculation.  Taking first the results for the Daejeon16
interaction [Fig.~\ref{fig:mirror-asymmetry-ratio-scan}(a)], the deviations from
unity found in the present calculations, at $\lesssim 5\%$, might seem modest
from a practical perspective, in terms of the precision we may desire in extracting the quadrupole moment ratio.  However,
the calculated ratio shows little indication of convergence with respect to
$\Nmax$.  While the $\hw$-dependence of the curves is modest, with a gentle peak
near $\hw=10\,\MeV$, curves for successive $\Nmax$ are nearly equidistant,
giving larger values of the ratio for increasing $\Nmax$.  Assuming this trend
continues, the present NCCI calculations provide only a lower bound on the
deviation from mirror symmetry which would be obtained (in the full, untruncated
many-body problem) for the quadrupole moments, assuming the Daejeon16
interaction.

The interactions used in \textit{ab initio} calculations have an isospin violating portion which is either simplified outright (to
the Coulomb interaction) or subject to notable uncertainties in the nuclear
part~\cite{epelbaum2009:nuclear-forces}.  
Both the
Daejeon16 and JISP16 interactions are purely isoscalar, before inclusion of the
Coulomb interaction, which is thus the only source of isospin symmetry
violation.  (The proton-neutron mass difference is commonly neglected in NCCI
calculations~\cite{kamuntavicius1999:isoscalar-hamiltonians,gueorguiev2010:nuclear-mass-a-body-interaction-iwnt10,caprio2020:intrinsic}.)
In contrast, the LENPIC interaction explicitly includes isospin violation from
the strong interaction (\textit{e.g.}, Ref.~\cite{epelbaum2009:nuclear-forces}).
It is not \textit{a priori} obvious how this might influence the deviations from
mirror symmetry in the quadrupole moment.

The violation of mirror symmetry in the $A=7$ pair is compared across
interactions in Fig.~\ref{fig:mirror-asymmetry-ratio-scan}~(top).  The scale of
the calculated deviations, for any given $\Nmax$ and $\hw$, is found to be
comparable across the Daejeon16 [Fig.~\ref{fig:mirror-asymmetry-ratio-scan}(a)]
and JISP16 [Fig.~\ref{fig:mirror-asymmetry-ratio-scan}(b)] interactions, not
surprisingly given their shared reliance on the Coulomb interaction as their
isospin symmetry breaking part, and marginally smaller for the LENPIC
interaction [Fig.~\ref{fig:mirror-asymmetry-ratio-scan}(c)].  However, there is
no obvious gross qualitative difference across interactions, and any detailed
quantitative comparison of these unconverged calculations is meaningless in
light of the different convergence rates for calculations with different
interactions.

\begin{figure*}
\begin{center}
\includegraphics[width=\ifproofpre{0.75}{1}\hsize]{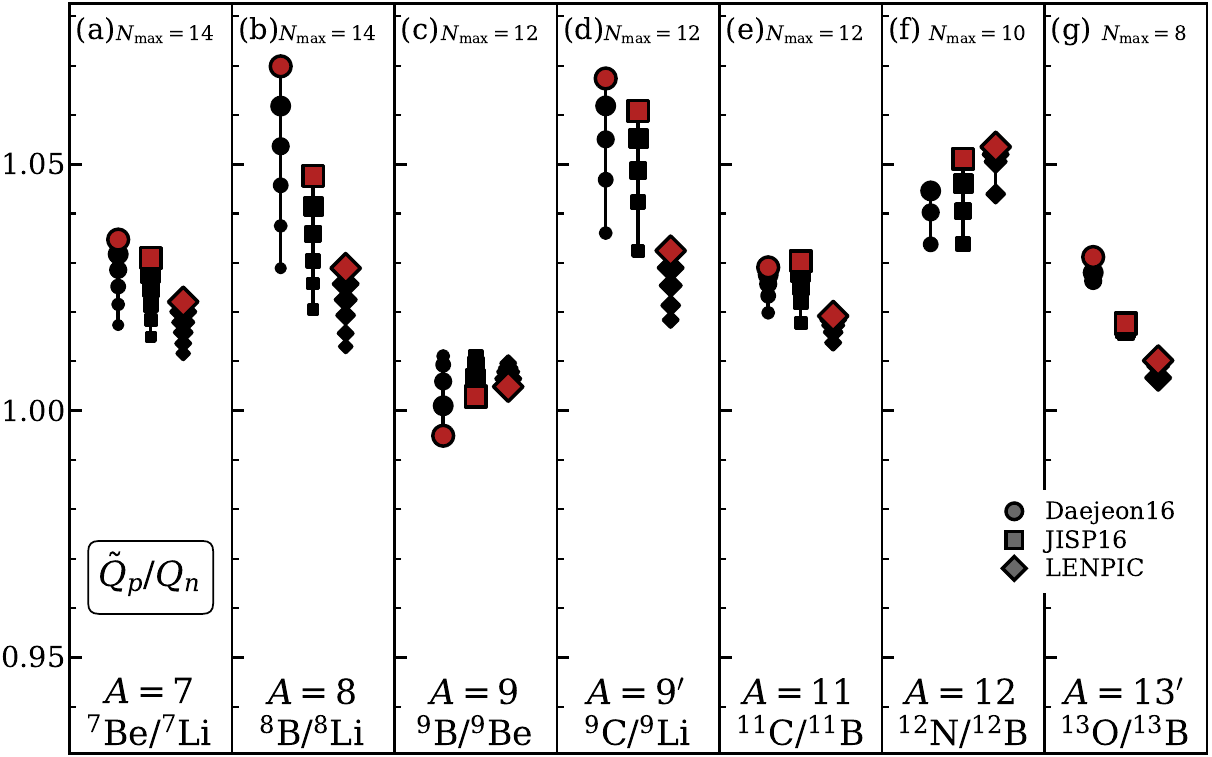}
\end{center}
\caption{Ratio of $Q_p$ in one member of mirror pair to $Q_n$ in the other,
  calculated with the Daejeon16~(circles), JISP16~(squares), and
  LENPIC~(diamonds) interactions at fixed $\hw$ ($15\,\MeV$, $20\,\MeV$, and
  $25\,\MeV$, respectively, for the three interactions).  Calculated values are
  shown for successive even values of $\Nmax$ (increasing symbol size), from
  $\Nmax=4$ to the maximum value for that mirror pair, indicated at top.
\label{fig:mirror-asymmetry-ratio-teardrop}
}
\end{figure*}

Proceeding to the remaining mirror pairs ($A>7$), we focus for purposes of
illustration on the results obtained with the Daejeon16 interaction.
(Comprehensive tabulations for all three interactions are again provided in the
Supplemental Material~\cite{supplemental-material}.)  The convergence properties
of the calculated $\tilde{Q}_p/Q_n$ are shown in
Fig.~\ref{fig:mirror-asymmetry-ratio-scan}~(bottom).  As in the $A=7$ example
above, we again take the ``nuclide of interest'' for calculating $Q_n/Q_p$,
in~(\ref{eqn:ratio-relation}), to be the neutron-rich ($Z<N$) member of the
mirror pair, and the ``mirror nuclide'' providing the ``proxy'' value
$\tilde{Q}_p$ to be the proton-rich ($Z>N$) member.  To provide a global
comparison across interactions, we also present the results for all three
interactions as functions of $\Nmax$ for fixed $\hw$ in
Fig.~\ref{fig:mirror-asymmetry-ratio-teardrop} (that is, in the same spirit as
Fig.~\ref{fig:mirror-ratio-teardrop} above).

For all mirror pairs, the calculated deviations of $\tilde{Q}_p/Q_n$ from unity,
as seen in Figs.~\ref{fig:mirror-asymmetry-ratio-scan}
or~\ref{fig:mirror-asymmetry-ratio-teardrop}, may again seem modest, at $< 10\%$.  But again
nonconvergence with respect to $\Nmax$ is the norm.  Following the same order of
discussion as in Sec.~\ref{sec:ratio}, let us proceed first through the odd-mass
nuclides, then return to the odd-odd nuclides.

For $A=9$ [Figs.~\ref{fig:mirror-asymmetry-ratio-scan}(e)
  and~\ref{fig:mirror-asymmetry-ratio-teardrop}(c)], the ratio is initially
greater than unity but decreases below unity for large $\Nmax$, and the changes in fact
are becoming larger for successive $\Nmax$.  While there is a sharp downturn in
the curves for $\hw\lesssim 10$, recall from Sec.~\ref{sec:ratio} that the
calculations for $\hw\lesssim 10\,\MeV$ for the Daejeon16 interactions may be of
limited relevance to the physical ground state.

For $A=9'$ [Figs.~\ref{fig:mirror-asymmetry-ratio-scan}(f)
  and~\ref{fig:mirror-asymmetry-ratio-teardrop}(d)], the spacing between
successive $\Nmax$ curves systematically decreases with increasing $\Nmax$,
hinting at eventual convergence, perhaps with a $\lesssim10\%$ violation of
mirror symmetry.  This case is of special interest in the context of isospin
symmetry, since these $\isotope[9]{Li}$ and $\isotope[9]{C}$ ground states
bookend a $T=3/2$ quartet of isobaric analog states noted for apparent violation
of the isobaric multiplet mass
equation~\cite{brodeur2012:a9-imme-shell,lam2013:imme-alt71}.  (The excited
analog states in $\isotope[9]{Be}$ and $\isotope[9]{B}$ are above the
single-nucleon separation threshold, and thus are subject to the Thomas-Ehrman
effect~\cite{ehrman1951:13c-13n-displacement,thomas1952:13c-13n-levels}.)

For $A=11$ [Figs.~\ref{fig:mirror-asymmetry-ratio-scan}(g)
  and~\ref{fig:mirror-asymmetry-ratio-teardrop}(e)], while the calculated
$Q_p(\isotope[11]{C})/Q_n(\isotope[11]{B})$ changes comparatively little with
$\Nmax$, and the curves in Fig.~\ref{fig:mirror-asymmetry-ratio-scan}(g) might
superficially be taken to suggest an $\sim3\%$ deviation from mirror symmetry,
closer inspection shows that the calculated ratio still increases steadily with
$\Nmax$.  There is thus no clear sign of convergence.

For the $A=13$ pair [Figs.~\ref{fig:mirror-asymmetry-ratio-scan}(i)
  and~\ref{fig:mirror-asymmetry-ratio-teardrop}(g)], recall from
Sec.~\ref{sec:ratio} that the convergence behavior of $Q_p(\isotope[13]{O})$,
and thus of the ratio, is anomalous [Fig.~\ref{fig:mirror-scan-agt7}(i)], at
least partly attributable to the semi-magic nature of these nuclei.  It is
therefore not obvious how to interpret the convergence behavior of the deviation
from mirror symmetry.

Then, returning to the odd-odd nuclides, for $A=8$
[Figs.~\ref{fig:mirror-asymmetry-ratio-scan}(d)
  and~\ref{fig:mirror-asymmetry-ratio-teardrop}(b)], the ratio continues to
march upwards, with approximately constant spacing between calculations for
successive $\Nmax$.  For $A=12$ [Figs.~\ref{fig:mirror-asymmetry-ratio-scan}(h)
  and~\ref{fig:mirror-asymmetry-ratio-teardrop}(f)], the spacing is gradually
decreasing, but the ultimate value to which this ratio will converge can only be
estimated as likely giving a $<10\%$ deviation from mirror symmetry.

It is interesting to compare these (unconverged) estimates of the deviation from
mirror symmetry in the quadrupole moment with those found in the GFMC
calculations~\cite{pastore2013:qmc-em-alt9}.  In
Ref.~\cite{pastore2013:qmc-em-alt9}, both $Q_p$ and $Q_n$ were calculated
independently for each member of the $A=8$ and $9'$ mirror pairs (see note to
Table~\ref{tab:expt-gfmc} above), yielding
$Q_p(\isotope[8]{B})/Q_n(\isotope[8]{Li})=0.91(7)$ and
$Q_p(\isotope[9]{C})/Q_n(\isotope[9]{Li})=1.11(11)$.  While both results are
consistent with unity (or nearly so) within GFMC statistical uncertainties,
intriguingly, the sense of the deviation differs from the present results for
$A=8$ [compare Fig.~\ref{fig:mirror-asymmetry-ratio-teardrop}(b)] but agrees for
$A=9'$ [compare Fig.~\ref{fig:mirror-asymmetry-ratio-teardrop}(d)].  The isospin
violation in the wave functions provided by the GFMC calculations is severely
restricted, as the GFMC propagator only takes the Coulomb interaction into
account approximately (as an effective isoscalar Coulomb operator with a
$T_z$-dependent normalization) and replaces the isospin-breaking AV18
interaction with the simpler isoscalar AV8$'$
interaction~\cite{pudliner1997:qmc-aleq7}.

In summary, although the isospin quantum number in the \textit{ab initio} NCCI
calculations is quite ``good'', we should not be lulled by this observation into assuming
the validity of mirror symmetry for observables in general, and for the quadrupole
moments in particular.  The traditional use of the mirror quadrupole moment as
a proxy for $Q_n$ must be treated with caution.  The present calculations
do not provide a firm estimate for the error thereby incurred: both since
convergence is slow in the many-body calculation for the relevant observables
(Fig.~\ref{fig:mirror-asymmetry-ratio-scan}), but also potentially due to
limitations in the isospin breaking contributions in the underlying
interactions.  Nonetheless, the calculations do suggest deviations of at least
$5$--$10\%$, and potentially much larger.

%% file: emratio-discussion.tex
\section{Interpretation of theoretical results}
\label{sec:discussion}

\begin{figure*}
\begin{center}
\includegraphics[width=\ifproofpre{0.75}{1}\hsize]{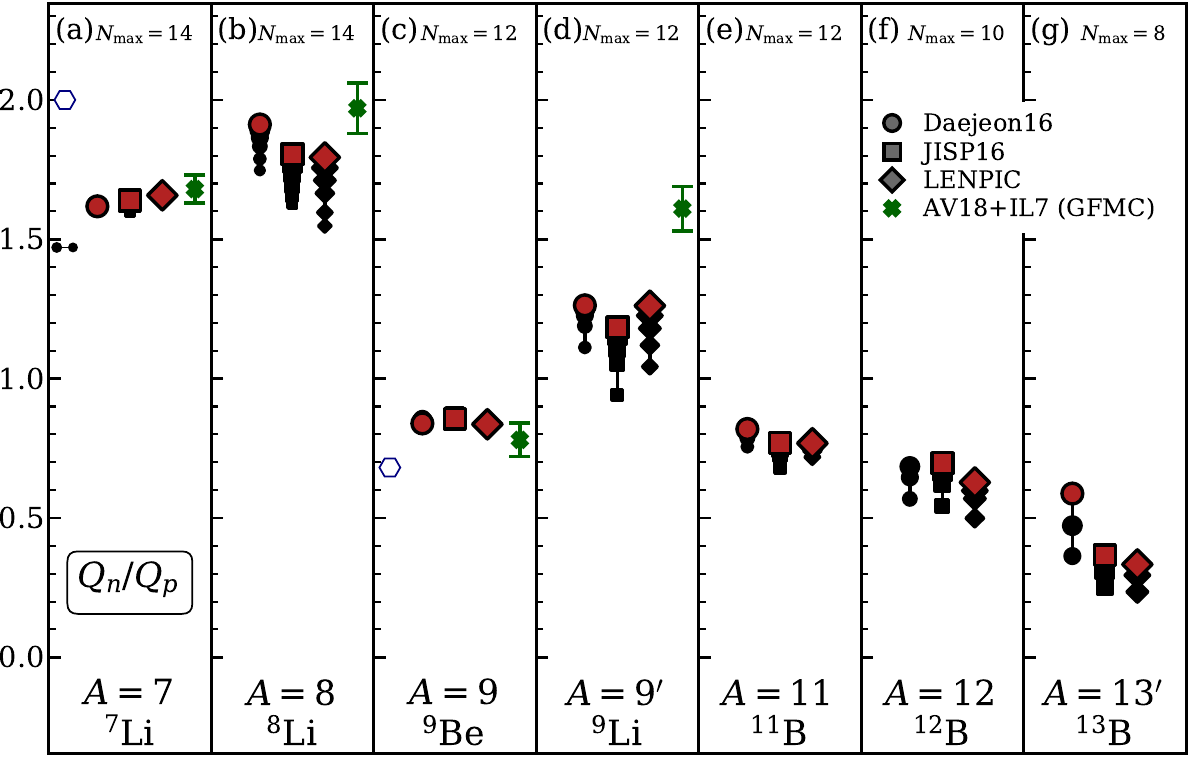}
\end{center}
\caption{Ratios $Q_n/Q_p$, for the neutron-rich members of the $p$-shell mirror
  pairs, calculated with the Daejeon16~(circles), JISP16~(squares), and
  LENPIC~(diamonds) interactions at fixed $\hw$ ($15\,\MeV$, $20\,\MeV$, and
  $25\,\MeV$, respectively, for the three interactions).  Calculated values are
  shown for successive even values of $\Nmax$ (increasing symbol size), from
  $\Nmax=4$ to the maximum value for that mirror pair, indicated at top.  Also
  shown (see Table~\ref{tab:expt-gfmc}) are the GFMC AV18+IL7
  predictions~\cite{pastore2013:qmc-em-alt9} (crosses), along with a schematic
  cluster model estimate (see text) for $\isotope[7]{Li}$ (dumbbell) and
  Elliott-Wilsdon $\grpsu{3}$ rotational model estimates for $\isotope[7]{Li}$
  and $\isotope[9]{Be}$ (open hexagons).
  \label{fig:neutron-proton-ratio-teardrop}
}
\end{figure*}

To extract information on the proton-neutron structure from calculations, of
course, it is not necessary to use the mirror ratio (Sec.~\ref{sec:ratio}) as a
proxy for the ratio of neutron and proton quadrupole moments.  This was simply
imposed by experimental necessity.  Rather, we may directly calculate $Q_n/Q_p$
from the wave function for a given nuclide.  Such results are shown for
reference for the neutron rich member of each mirror pair in
Fig.~\ref{fig:neutron-proton-ratio-teardrop}, from the present NCCI wave
functions, alongside analogous results (for $A\leq9$) from the GFMC calculations
of Ref.~\cite{pastore2013:qmc-em-alt9}.  Qualitatively there is little
difference from the mirror ratio results shown in
Fig.~\ref{fig:mirror-ratio-teardrop}, to which these results are related by the
mirror asymmetry ratios of Fig.~\ref{fig:mirror-asymmetry-ratio-teardrop} (which
are admittedly unconverged, but remain numerically close to unity for the
present calculations).

If qualitative understanding of the nuclear structure is sought, some further
conceptual framework (or model) is needed to interpret these raw computed
values.  Here, after a few general initial comments, we simply note some
possible fruitful avenues, involving clustering and dynamical symmetry.

A naive baseline estimate for $Q_n/Q_p$ may be made in the spirit of the
collective liquid drop interpretations of heavier nuclei, by taking the nucleus
as a homogeneously charged matter distribution.  In this case, the quadrupole
operators for the protons and neutrons are identical except for overall
normalization, proportional to $Z/A$ for the protons~\cite{eisenberg1987:v1} and
$N/A$ for the neutrons.  We thus have $Q_n/Q_p=N/Z$, giving
$Q_n/Q_p=4/3\approx1.33$ for $\isotope[7]{Li}$, $Q_n/Q_p=5/3\approx1.67$ for
$\isotope[8]{Li}$, and so on, through $Q_n/Q_p=8/5\approx1.6$ for
$\isotope[13]{B}$.  While these estimates are not egregiously far off from the
calculations in Fig.~\ref{fig:neutron-proton-ratio-teardrop} (or the
experimental mirror ratios in Fig.~\ref{fig:mirror-ratio-teardrop}), except in
the ill-behaved case of $A=13'$, neither do their variations from nuclide to
nuclide particularly match or illuminate the trends in the ratios as one moves
across Fig.~\ref{fig:neutron-proton-ratio-teardrop}.

The viewpoint that collectivity arises primarily in the valence shell
might suggest rather that the numbers of \textit{valence} nucleons of each
species, \textit{i.e.}, $N_p$ for the protons and $N_n$ for the
neutrons~\cite{casten2000:ns}, are more relevant than the total numbers, $Z$ and
$N$.  However, this raises the thorny issue of the appropriate effective
charges $e_p$ and $e_n$ to apply in an estimate $Q_n/Q_p=(e_nN_n)/(e_pN_p)$.

Indeed, the contrasting traditional model framework would be the shell model,
and certainly valence shell calculations in the $p$ shell can bear striking
similarity to \textit{ab initio} results~\cite{johnson2015:spin-orbit} (see
Fig.~13 of Ref.~\cite{caprio2015:berotor-ijmpe}).  Detailed shell model
studies~\cite{smirnova2003:mirror-gt-second-class} of the ground state
quadrupole moments for several of the present nuclei, based on empirical
isospin-nonconserving shell-model
interactions~\cite{ormand1989:inc-interaction-shell}, highlight the dependence
of the calculated quadrupole moments in detail not only upon the effective
charges but also upon the choice of shell model radial wave function (harmonic
oscillator \textit{vs.}\ Woods-Saxon).

Given the prominent role of clusterization in the structure of $p$-shell
nuclides (\textit{e.g.}, Refs.~\cite{freer2007:cluster-structures}), we may
naturally seek to use \textit{ab initio} calculated quadrupole moment ratios to
test or provide constraints on interpretations based on clustering and the
associated cluster molecular orbitals.  In a clustering description, the naive
assumption of a homogeneous charge distribution is manifestly broken, except in
a few special cases (\textit{e.g.}, pure $\alpha$-cluster nuclei).  For
instance, in a cluster molecular orbital picture, the neutron-rich
$\isotope{Be}$ isotopes consist of a $2\alpha$ dimer plus additional ``valence''
neutrons in molecular
orbitals~\cite{hiura1963:alpha-model-9be,*hiura1964:alpha-model-9be-ERRATUM,okabe1979:9be-molecular-part-1,*okabe1979:9be-molecular-part-2,seya1981:molecular-orbital,vonoertzen1996:be-molecular,vonoertzen1997:be-alpha-rotational,itagaki2000:10be-molecular-orbital,itagaki2000:be-molecular-orbital-spin-orbit,freer2007:cluster-structures,dellarocca2017:cluster-orbitals,dellarocca2018:cluster-shell-model-part1-9be-9b}.
A microscopic justification for this description is obtained from
antisymmetrized molecular dynamics
(AMD)~\cite{kanadaenyo1997:c-amd-pn-decoupling,kanadaenyo1999:10be-amd,suhara2010:amd-deformation,kanadaenyo2012:amd-cluster},
fermionic molecular dynamics (FMD)~\cite{zakova2010:10be-fmd}, or \textit{ab
  initio} resonating group method
(RGM)~\cite{kravvaris2017:ab-initio-cluster-8be-10be-12c} calculations of the
$\isotope{Be}$ isotopes.  The neutron quadrupole moment is sensitive to whether the
neutrons occupy $\pi$ (``equatorial'') or $\sigma$ (``polar'') molecular
orbitals, while the proton quadrupole moment is largely sensitive only to the
inter-$\alpha$ separation.

Here it is interesting to note what we would expect from a schematic ``ball and
stick'' molecular model for $\isotope[7]{Li}$.  That is, suppose we take the
ground state of $\isotope[7]{Li}$ to have an $\alpha+t$ cluster structure, and
then we reduce this description to the simple, naive limit of two point-like
clusters, separated by a fixed, finite distance.  The resulting structure would
be an axially-symmetric rigid rotor.  By the usual proportionality between
laboratory-frame and intrinsic quadrupole moments for an axially-symmetric
rotor~\cite{rowe2010:collective-motion}, the ratio of the neutron and proton
quadrupole moments in the laboratory frame would be identical to their ratio in the
rotational intrinsic (principal axes) frame, that is, computed with respect to
the molecular symmetry axis.  This yields\footnote{The proton and neutron
  quadrupole moments must be calculated relative to the common center of mass,
  which lies $3/7$ of the way from the alpha to the triton, putting the clusters
  at coordinates $z_\alpha=-3/7$ and $z_t=+4/7$, respectively, along the
  symmetry axis (in units of an arbitrary separation $\ell$, which is irrelevant
  to the final ratio).  Then, $Q_p=2z_\alpha^2+z_t^2$, and
  $Q_n=2z_\alpha^2+2z_t^2$, giving $Q_n/Q_p=50/34\approx1.47$.} an estimate of
$Q_n/Q_p$ just shy of $1.5$, as indicated in
Fig.~\ref{fig:neutron-proton-ratio-teardrop}(a) (dumbbell shape).  Clearly, this
picture ignores essential features of any more realistic cluster picture, such
as the finite size of the clusters, which is comparable in scale to their
separation, the consequent modification of the clusters by Fermi exclusion
effects, as well as other polarization effects on the internal structure of the
clusters, and zero-point oscillations in their separation.  Yet the \textit{ab
  initio} predictions, which are robustly consistent with each other in the
range $Q_p/Q_n\approx1.6$--$1.7$, provide a ratio only $10$--$15\%$ larger than
this schematic estimate.

At the other end of the mass range, recall the ill-behaved convergence of the
proton quadrupole moment in $\isotope[13]{O}$
[Fig.~\ref{fig:mirror-scan-agt7}(i)].  This behavior, which now appears in the
neutron quadrupole moment of $\isotope[13]{B}$, entering into the $Q_n/Q_p$
ratio of Fig.~\ref{fig:neutron-proton-ratio-teardrop}(g), takes on new
significance in a cluster molecular orbital interpretation, where
$\isotope[13]{B}$ is described as $2\alpha+4n+p$.  Here, it has been
proposed~\cite{okabe:cited} that the large ground state (proton) quadrupole
moment reflects the result of mixing between two low-lying $3/2^-$ cluster
configurations, obtained by coupling a proton in a $\pi$ orbital (with $K=3/2$)
to the two low-lying $0^+$ states of $\isotope[12]{Be}$.  These
$\isotope[12]{Be}$ states, in turn, differ not only in inter-$\alpha$
separation, which affects the proton quadrupole moment, but also in whether the
molecular orbitals occupied by the neutrons are $\pi^4$ (a low-deformation
configuration) or $\pi^2\sigma^2$ (a high-deformation
configuration)~\cite{itagaki2000:be-molecular-orbital-spin-orbit} (see also
Fig.~5 of Ref.~\cite{kanadaenyo2012:amd-cluster}), which may be expected to
dramatically affect the neutron quadrupole moment.

Thus, in a clustering picture, the ground state $Q_p$ and $Q_n$ in
$\isotope[13]{B}$ reflect the shape coexistence in the low-lying spectrum of
$\isotope[12]{Be}$ (\textit{e.g.}, Ref.~\cite{shimoura2007:12be-lifetime}). Both
moments, and $Q_n$ in particular, should be sensitive to the resultant mixing of
configurations with significantly different deformations.  Meanwhile,
in NCCI calculations, the relative energy for the $\isotope[12]{Be}$ $0^+$
states evolves rapidly with $\Nmax$ (\textit{e.g.}, Fig.~19 of
Ref.~\cite{maris2015:berotor2}), ostensibly making any such mixing highly
$\Nmax$ dependent.  Consistent with the proposed clustering picture of
$\isotope[13]{B}$, the NCCI calculations also yield a predominantly $2\hw$
excited $3/2^-$ state, which converges rapidly downward in energy towards the
predominantly $0\hw$ $3/2^-$ ground state, becoming the first excited state at
high $\Nmax$.

We may also seek to at least qualitatively understand the deviations from mirror
symmetry (Sec.~\ref{sec:symm}) in terms of cluster structure.  In nuclei where
prolate quadrupole deformation arises predominantly from the cluster dimer
structure, the additional Coulomb repulsion obtained in going from the
neutron-rich member of the mirror pair to the proton-rich member may be
expected to increase the inter-cluster separation, and thus the quadrupole
deformation.  Thus, \textit{e.g.}, in going from the neutrons in
$\isotope[7]{Li}$ to the protons in $\isotope[7]{Be}$
[Fig.~\ref{fig:mirror-asymmetry-ratio-teardrop}(a)], the increase in quadrupole
moment $Q_p$ for $\isotope[7]{Be}$ relative to $Q_n$ for $\isotope[7]{Li}$ would
be interpreted as arising from the increased repulsion and thus intercluster
separation between the alpha and helion clusters in $\isotope[7]{Be}$
($=\alpha+h$) as compared to the alpha and triton clusters in $\isotope[7]{Li}$
($=\alpha+t$).  (If variation in the internal structure of the clusters is also
considered, then the extra Coulomb repulsion would also be expected to enhance
the proton quadrupole moment by increasing the polarization of the clusters,
such that the protons are further displaced towards the termini of the
molecule.)

In contrast, if $\isotope[9]{Be}$ and $\isotope[9]{B}$
[Fig.~\ref{fig:mirror-asymmetry-ratio-teardrop}(c)] are taken as $2\alpha+n$ and
$2\alpha+p$, respectively, the valence nucleon is expected to be in an
equatorial $\pi$
orbital~\cite{hiura1963:alpha-model-9be,*hiura1964:alpha-model-9be-ERRATUM,vonoertzen1996:be-molecular,freer2007:cluster-structures,dellarocca2018:cluster-shell-model-part1-9be-9b}
(see also Fig.~5 of Ref.~\cite{maris2012:mfdn-ccp11} for \textit{ab initio}
predictions).  This nucleon thus gives an oblate contribution to the
deformation, serving to reduce the overall positive intrinsic quadrupole moment
calculated with respect to the symmetry axis of the $2\alpha$ dimer.  An
increase in the size of this orbital, induced by Coulomb repulsion, would
therefore tend to reduce $Q_p$ in $\isotope[9]{B}$ relative to $Q_n$ in
$\isotope[9]{Be}$, an effect which would be in competition with any concomitant
increase in the inter-$\alpha$ spacing in $\isotope[9]{B}$ relative to
$\isotope[9]{Be}$.

Symmetry-based descriptions provide an alternative route by which we may seek to
understand the proton-neutron quadrupole structure revealed in \textit{ab
  initio} calculations.  The ground state wave functions of a variety of
$p$-shell nuclides, in \textit{ab initio} NCCI calculations, have been found to
be remarkably well-described by a dominant contribution exhibiting Elliott
$\grpsu{3}$
symmetry~\cite{dytrych2013:su3ncsm,launey2016:sa-nscm,kravvaris2018:be-clustering-sotancp4,mccoy2018:diss,mccoy2018:spncci-busteni17,mccoy2020:spfamilies,zbikowski2021:beyond-elliott}.
Elliott's $\grpsu{3}$
group~\cite{elliott1958:su3-part1,*elliott1958:su3-part2,*elliott1963:su3-part3,*elliott1968:su3-part4,harvey1968:su3-shell}
has as its generators the components of a quadrupole tensor $\calQ_2$ (this is
Elliott's restricted quadrupole operator, which conserves oscillator quanta) and
the orbital angular momentum vector $L_1$ (\textit{e.g.}, Appendix~A.4 of
Ref.~\cite{caprio2020:intrinsic}).  Elliott's $\grpsu{3}$ thus provides a
symmetry-based description for the correlations which give rise to quadrupole
deformation and rotation.

For the $p$-shell nuclides, Elliott's $\grpsu{3}$ model provides definite
predictions for the quadrupole moment ratio $Q_n/Q_p$.
In Elliot's $\grpsu{3}$ model, the ground state is expected to come from the
``leading'' $\grpsu{3}$ irreducible representation (irrep) in the $0\hw$ space.
The states constituting this irrep are the most deformed in the $0\hw$ space,
and thus most bound by Elliott's $-\calQ\cdot\calQ$ schematic
Hamiltonian~\cite{harvey1968:su3-shell}.  If we work in a proton-neutron
$\grpsu{3}$ scheme, described by the group chain
$\grpsu[p]{3}\times\grpsu[n]{3}\supset\grpsu{3}$, then the proton and neutron
quadrupole operators are also generators.  This permits simple analytic
calculations for the proton and neutron quadrupole moments, to within an
arbitrary overall scale (depending on the oscillator length of
Sec.~\ref{sec:ratio:a7}), which cancels in their ratio.

The ground-state $\grpsu{3}$ structure for $\isotope[7]{Li}$ (or, equivalently,
its mirror nuclide $\isotope[7]{Be}$) has been explored in \textit{ab initio}
calculations~\cite{mccoy2018:diss,mccoy2020:spfamilies,caprio2020:bebands},
showing that the $3/2^-$ ground state indeed comes predominantly from the
leading irrep in the $\isotope[7]{Li}$ $0\hw$ space, which has $\grpsu{3}$
quantum numbers $(\lambda,\mu)=(3,0)$ and total spin $S=1/2$.  The ground state
in this description is uniquely defined, as a pure $LS$-coupling scheme state
in which orbital angular momentum $L=1$ combines with the spin to give $J=3/2$
(see Fig.~1 of Ref.~\cite{mccoy2020:spfamilies}).  The resulting $\grpsu{3}$
estimate $Q_n/Q_p=2$, shown in Fig.~\ref{fig:mirror-asymmetry-ratio-teardrop}(a)
(open hexagon), errs in the opposite direction from the naive cluster estimate
above.

The expected ground state $\grpsu{3}$ structure for $\isotope[9]{Be}$ has also
been discussed
extensively~\cite{millener2001:light-nuclei,millener2007:p-shell-hypernuclei,caprio2020:bebands}.
In the leading $\grpsu{3}$ irrep, which is $(3,1)$, with $S=1/2$, a $J=3/2$ state
can be constructed two ways, by combining either $L=1$ or $L=2$ with the spin to
give $J=3/2$.  The physical ground state may be expected to be some mixture of
these.  Although the $L=1$ state has lower rotational energy, in the
Elliott-Wilsdon rotational
model~\cite{elliott1968:su3-part4,harvey1968:su3-shell} the spin-orbit
interaction mixes these states to give a $K=3/2$ rotational bandhead state
$\tket{K=3/2;J=3/2}=\sqrt{21/26}\tket{L=1;J=3/2}-\sqrt{5/26}\tket{L=2;J=3/2}$~\cite{millener2001:light-nuclei}.
The resulting $\grpsu{3}$ estimate $Q_n/Q_p\approx0.68$
[Fig.~\ref{fig:mirror-asymmetry-ratio-teardrop}(c)] lies only marginally below
the \textit{ab initio} predictions.

%% file: emratio-concl.tex
\section{Conclusion}
\label{sec:conclusion}

In summary, meaningful predictions of electric quadrupole moment ratios can be
made, in \textit{ab initio} NCCI calculations, even though the moments themselves
are not individually converged.  This observation applies more generally to
ratios of electromagnetic matrix elements involving states with similar
convergence properties, arising from structural similarities. Examples include
(ratios of) transition strengths within a rotational
band~\cite{maris2015:berotor2,caprio2020:bebands}, or between rotational bands
with related structures~\cite{caprio2019:bebands-sdanca19}, or across mirror
transitions~\cite{henderson2019:7be-coulex}.

In particular, ratios of electric quadrupole moments in mirror nuclei provide an
observable which can be both, on one hand, precisely measured experimentally
and, on the other hand, well-converged in NCCI calculations, thereby providing
stringent tests of the theoretical framework.  Where the experimental ratio is
known, and where a robust prediction can be meaningfully extracted from the NCCI
calculations, we find generally good agreement (Fig.~\ref{fig:mirror-ratio-teardrop}).
Alternatively, where the moment of only one member of the mirror pair is known,
NCCI calculations can be taken to provide predictive power for the other unknown
moment.

However, it must be emphasized that precision \textit{ab initio} predictions of
electric quadrupole moment ratios are not always possible.  A robust prediction
for the ratio may be understood as arising when the incomplete convergence of
the quadrupole moments themselves is simply a systematic effect of the basis
truncation, to be cancelled out between the moments, but not if the moments are
exquisitely sensitive to fine details of the many-body calculation, \textit{e.g.},
delicate mixing of competing low-lying configurations.

By isospin mirror symmetry, the ratio of electric quadrupole moments across
mirror nuclei is closely related, though not strictly equivalent
(Fig.~\ref{fig:mirror-asymmetry-ratio-teardrop}), to the neutron/proton
quadrupole moment ratio within a single nuclide.  Although the neutron
quadrupole moment itself is not directly accessible experimentally, \textit{ab
  initio} calculations can provide robust predictions for the neutron/proton
quadrupole moment ratio (Fig.~\ref{fig:neutron-proton-ratio-teardrop}), thereby
giving insight into the isovector aspects of the quadrupole deformation.  In the
limit of axially-symmetric adiabatic rotation, this ratio measures the relative
contributions of the neutrons and protons to the deformation.  More generally,
it is subject to interpretation through effective descriptions of the
proton-neutron structure of the nucleus, \textit{e.g.}, involving simpler
degrees of freedom, as in clustering, or correlations imposed by symmetry, as in
the Elliott $\grpsu{3}$ picture.

%% file: emratio.bbl
%